\documentclass[11pt]{article}
\usepackage{jcapmod}

\usepackage{booktabs}
\usepackage[english]{babel}
\usepackage{amsmath,amssymb,amsbsy,amstext, amsthm, simplewick, amsfonts}
\usepackage{hyperref}
\usepackage{graphicx}
\usepackage[small]{caption}
\usepackage{siunitx}
\usepackage{upgreek}
\usepackage{framed}
\usepackage{wrapfig}
\usepackage{multirow}
\usepackage{subfig}
\usepackage{bbm}
\usepackage[svgnames,dvipsnames,x11names]{xcolor}
\usepackage[utf8x]{inputenc}
\usepackage{selinput}
\usepackage{bm}
\usepackage{float}
\usepackage{yfonts}
\usepackage[bottom]{footmisc}

\def\d{{\rm d}}

\def\k{{\boldsymbol k}}

\def\q{{\boldsymbol q}}

\def\x{{\boldsymbol x}}

\def\0{\boldsymbol{0}}

\def\L{{\cal L}}

\def\O{{\cal O}}

\def\fnl{{f_{\rm NL}}}
\def\hfnl{{\hat f_{\rm NL}}}

\def\nn{\nonumber\\}

\def\hs{\hskip 1pt}

\def\bea{\begin{eqnarray}}
\def\eea{\end{eqnarray}}
\def\be{\begin{equation}}
\def\ee{\end{equation}}

\usepackage[framemethod=default]{mdframed}
\newmdenv[skipabove=7pt,
skipbelow=7pt,
rightline=false,
leftline=false,
topline=false,
bottomline=false,
backgroundcolor=gray!10,
linecolor=gray,
innerleftmargin=5pt,
innerrightmargin=5pt,
innertopmargin=5pt,
innerbottommargin=3pt,
leftmargin=0cm,
rightmargin=0cm,
linewidth=4pt]{eBox}

\begin{document}
\begin{titlepage}

\setcounter{page}{1} \baselineskip=15.5pt \thispagestyle{empty}

\bigskip\

\vspace{1cm}

\begin{center}
{\fontsize{19}{0}\selectfont  \sffamily \bfseries Capturing Non-Gaussianity of the  \\[12pt] Large-Scale~Structure~with~Weighted~Skew-Spectra
}
\vspace{0.6cm}
\end{center}

\begin{center}
{\fontsize{13}{30}\selectfont  Azadeh Moradinezhad Dizgah,$^{1,2}$  Hayden Lee,$^{2}$ Marcel Schmittfull$^{3}$, and Cora Dvorkin$^{2}$} 
\end{center}

\begin{center}

\vskip 15pt
\textsl{$^1$ Department of Theoretical Physics and Center for Astroparticle Physics (CAP), \\
University of Geneva, 24 quai E. Ansermet, Geneva, Switzerland }   
\vskip 8pt
\textsl{$^2$ Department of Physics, Harvard University, 17 Oxford Street, Cambridge, MA 02138, USA}  
\vskip 8pt
\textsl{$^3$School of Natural Sciences, Institute for Advanced Study,  Princeton, NJ 08540, USA }
\end{center}

\vspace{1.2cm}
\noindent {\sffamily \bfseries Abstract\vspace{.2cm}}\\
\noindent The forthcoming generation of wide-field galaxy surveys will probe larger volumes and galaxy densities, thus allowing for a much larger signal-to-noise ratio for higher-order clustering statistics, in particular the galaxy bispectrum. Extracting this information, however, is more challenging than using the power spectrum due to more complex theoretical modeling, as well as significant computational cost of evaluating the bispectrum signal and the error budget. To overcome these challenges, several proxy statistics have been proposed in the literature, which partially or fully capture the information in the bispectrum, while being computationally less expensive than the bispectrum. One such statistics are {\it weighted skew-spectra}, which are  cross-spectra of the density field and appropriately weighted quadratic fields. Using Fisher forecasts, we show that the information in these skew-spectra is equivalent to that in the bispectrum for parameters that appear as amplitudes in the bispectrum model, such as galaxy bias parameters or the amplitude of primordial non-Gaussianity. We consider three shapes of the primordial bispectrum: local, equilateral and that due to massive particles with spin two during inflation. To obtain constraints that match those from a measurement of the full bispectrum, we find that it is crucial to account for the full covariance matrix of the skew-spectra.

\vspace{1.in}
\noindent \line(1,0){250} \\
\noindent Email:{ \href{Azadeh.MoradienzhadDizgah@unige.ch}{Azadeh.MoradienzhadDizgah@unige.ch}} 
\end{titlepage}

\hrule
\tableofcontents
\vspace{.6cm}
\hrule
\vspace{.4cm}


\section{Introduction}

High-precision measurements of the cosmic microwave background (CMB) anisotropies have so far played a central role in shaping our understanding of the origin of the Universe and its evolution~\cite{Aghanim:2018eyx}. The next generation of CMB observations will continue to provide invaluable cosmological information via higher-precision measurements of the CMB primary polarization anisotropies, and the CMB secondary anisotropies, such as CMB lensing~\cite{Ade:2018sbj,Abazajian:2019eic}. This will be complemented by the massive amount of high-precision data from upcoming wide-field galaxy surveys, such as DESI \cite{Aghamousa:2016zmz}, EUCLID \cite{Amendola:2016saw}, SPHEREx \cite{Dore:2014cca}, and LSST \cite{Abell:2009aa}, measuring the shape and clustering statistics of  galaxies and quasars over large comoving volumes and a wide range of redshifts.

To date, most of the cosmological information from the large-scale structure (LSS) of the universe has been extracted from the 2-point clustering statistics, such as the power spectrum.
Given the inherent nonlinearity of the LSS, it is well known that there is a wealth of information in higher-order statistics~\cite{Scoccimarro:2000sn,Sefusatti:2006pa}. Accounting for this will be essential in recovering information not present in the power spectrum, e.g.~the imprint of primordial bispectrum beyond the local shape. Moreover, higher-point functions can be used to measure nonlinear bias parameters, reduce degeneracies that are present at the level of the power spectrum, and test the nonlinear modeling of the power spectrum. Future LSS surveys promise larger volumes and galaxy densities, thus allowing for a much larger signal-to-noise ratio for higher-order clustering statistics, in particular the bispectrum. Extracting the information from the bispectrum, however, is more challenging than it is from the power spectrum, due to complexities in theoretical modeling of the bispectrum, as well as significant computational cost of measuring the signal and determining the error budget \cite{Colavincenzo:2018cgf,Oddo:2019run}. Alternatively, to bypass some of these challenges, several proxy statistics have been proposed in the literature that attempt to capture the information of the LSS bispectrum on parameters of interest, and are simpler and computationally less expensive than the bispectrum \cite{Fergusson:2010ia,Regan:2011zq,Schmittfull:2012hq,Schmittfull:2014tca,Chiang:2015pwa,Eggemeier:2015ifa,Eggemeier:2016asq,Wolstenhulme:2014cla}. 

In practice, different proxy statistics compress the information in the bispectrum via some weighted averaging of the bispectrum signal. Depending on how this averaging is performed, the proxies may have less information than the bispectrum, either due to compression or due to being sensitive to only certain triangle configurations. It is therefore important to construct the proxies such that they optimally capture the information of the bispectrum on the parameters that one is after. An extensive comparison of information content of three of these proxies and constraints on cosmological parameters, integrated bispectrum, line bispectrum and model bispectrum has been performed in Ref.~\cite{Byun:2017fkz}.

In this paper, we focus on the LSS bispectrum estimator introduced in Ref.~\cite{Schmittfull:2014tca}, which we will refer to as {\it weighted skew-spectra} to connect with similar previously introduced estimators for the CMB and LSS~\cite{Cooray:2001ps,Munshi:2009ik,Munshi:2010df,Pratten:2011kh}. The skew-spectra are constructed so that their combined information on amplitude-like parameters, such as bias parameters and the amplitude of primordial non-Gaussianity (PNG), is equivalent to that of the bispectrum.
Computationally, evaluation of the skew-spectra is equivalent to power spectrum estimation. Focusing on Gaussian initial conditions, in Ref.~\cite{Schmittfull:2014tca} these estimators and their covariances were measured on $N$-body simulations and compared with theoretical predictions. Furthermore, the estimators were used to measure halo bias parameters and corrections to the shot noise of the bispectrum. An explicit comparison with bispectrum constraints, however, was not performed. Also, although the corresponding estimator for the contribution of the primordial bispectrum of the local-shape was introduced, no analysis was performed in that case. In this paper, we extend the work of Ref.~\cite{Schmittfull:2014tca} in several ways.
First, we compare the constraints on parameters expected from the skew-spectra with those from the bispectrum by performing a Fisher forecast. Through this comparison, we point out that some of the simplifying assumptions made in Ref.~\cite{Schmittfull:2014tca}, especially about the covariance, need to be corrected in order to capture the full information of the bispectrum. Furthermore, we also consider the contribution from primordial non-Gaussianity of local and equilateral shapes, as well as that due to the presence of massive particles with spin $s=2$ during inflation.

The rest of the paper is organized as follows. We start in Section~\ref{sec:cross_rev} by reviewing the construction of the skew-spectra as optimal estimators of the large-scale structure bispectrum. In Section~\ref{sec:cross_png}, we first discuss the separability of the template of the bispectrum due to the presence of massive particles with spin during inflation, and then proceed to construct the skew-spectra given this template, as well as local and equilateral primordial bispectra. In Section~\ref{sec:cross_grav}, we briefly review the results of Ref.~\cite{Schmittfull:2014tca} for matter and galaxy skew-spectra due to gravitational evolution, and in Section~\ref{sec:cross_shot}, we discuss the shot-noise contribution to the galaxy skew-spectrum. In Section~\ref{sec:covs}, we discuss the covariances of the skew-spectra and, in Section~\ref{sec:fisher}, we detail our forecasting methodology and survey design and present the results of our Fisher forecast. We draw our conclusions in Section~\ref{sec:con}. 
 
\section{Estimating the Bispectrum with Weighted Skew-Spectra}
\label{sec:cross_rev}
The standard method to constrain parameters with the bispectrum is as follows. First, the bispectrum is measured for various wavevector triangles, i.e.~triplets of 3D Fourier modes of the galaxy overdensity.
One then assumes a Gaussian likelihood for this measured bispectrum using a theoretical model for the expectation value of the bispectrum. This model depends on the parameters of interest, which commonly include cosmological parameters, galaxy bias and shot noise.
To obtain posteriors of these parameters, the likelihood, multiplied by priors on the parameters, is evaluated for many sample values of the parameters.
This general Markov chain Monte Carlo (MCMC) sampling method works for all parameters, no matter how complicated they enter the model.
However, this procedure requires measuring the bispectrum for a large number of Fourier triangles, ideally all triangles that can be formed, which is computationally slow and difficult to visualize.

To address this issue, we make the same assumptions as before, but we restrict ourselves to parameters that enter the model bispectrum linearly, noting that this includes a number of physically interesting examples such as galaxy bias parameters, the shot noise amplitude, the root mean square normalization of fluctuations $\sigma_8$, and primordial non-Gaussianity. The key consequence of this assumption is that the likelihood regarded as a function of these parameters has the form of a Gaussian, centered at the maximum-likelihood value for these parameters (with width given by the Fisher matrix).
Therefore, if we assume Gaussian priors for these parameters,\footnote{We make this assumption to simplify the discussion. For other types of priors, the resulting posterior in general would not be Gaussian, but may still be computable.} their posteriors will also be Gaussian, since the product of two Gaussian probability distributions is also Gaussian.
To compute this Gaussian posterior, all we need to compute from the data are therefore the maximum-likelihood estimates for the parameters.
The likelihood is then given by a Gaussian centered at those values, and can be multiplied by the prior to get the posterior, which will be identical to that obtained from MCMC sampling the parameters under our assumptions. This motivates finding a fast method to obtain the maximum likelihood parameter values following from the LSS bispectrum.

Let us start by considering just a single parameter, $f_{\rm NL}$, that enters the theoretical bispectrum model as an overall linear amplitude,
\be 
\langle\delta(\k_1)\delta(\k_2)\delta(\k_3)\rangle_{\rm th} =  (2\pi)^3 \delta_D(\k_1+\k_2+\k_3)f_{\rm NL} B_\delta^{\rm th}(\k_1,\k_2,\k_3)\, ,
\ee
where $\delta$ can be either matter or galaxy overdensity. The maximum likelihood estimator of this amplitude, $\hat f_{\rm NL}$, in the limit of weak non-Gaussianity is given by~\cite{Fergusson:2010ia} 
\begin{equation}
	\hat f_{\rm NL} =\frac{(2\pi)^3}{N_{\rm th}}\int\! \frac{\d^3k\hs\d^3q}{(2\pi)^6}  \frac{[\delta(\q)\delta(\k-\q)\delta(-\k)-3\langle\delta(\q)\delta(\k-\q)\rangle\delta(-\k)]B_\delta^{\rm th}(\k,\q,\k-\q)}{P_\delta(q)P_\delta(|\k-\q|) P_\delta(k)}\, ,
\label{eq:fnl_estimator}	
\end{equation}
where in writing the denominator, we have assumed that the covariance matrix is diagonal, so $P_\delta$ is the power spectrum of the observed density perturbation $\delta$, which includes shot noise for discrete tracers. Up to here the estimator in Eq.~\eqref{eq:fnl_estimator} is general and is valid in real and redshift space. The estimator has been normalized such that $\langle\hfnl\rangle=\fnl $, if the observed bispectrum is $B_\delta^{\rm obs} = f_{\rm NL} B_\delta^{\rm th}$.
Assuming statistical isotropy (i.e.~the bispectrum being only a function of three numbers), the normalization factor is therefore given by
\begin{align}
	N_{\rm th}=  \frac{V}{\pi}\int_{V_B}\d k_1\d k_2\d k_3\, \frac{k_1k_2k_3[B_\delta^{\rm th}(k_1,k_2,k_3)]^2}{P_\delta(k_1)P_\delta(k_2)P_\delta(k_3)}\, ,
\end{align}
where $V_B$ denotes the region over which the momenta satisfy the triangular conditions and $V=(2\pi)^3\delta_D(\0)$ is the survey volume. Hereafter we neglect the term linear in $\delta$ in Eq.~\eqref{eq:fnl_estimator}, since we are assuming the field to be statistically homogeneous (neglecting possible inhomogeneous noise). In writing the above expression for the estimator, it is assumed that we only account for the modes in the mildly nonlinear regime, where the bispectrum covariance matrix is nearly diagonal. The factors of the power spectra in the denominator therefore account for the inverse variance weighting of the observed density perturbation $\delta$. 

Suppose now that the theoretical bispectrum of interest can be expressed as a  sum of product separable terms of the form 
\begin{align}
	B_\delta^{\rm th}(\k_1,\k_2,\k_3)=f(\k_1)g(\k_2)h(\k_3)\, .
\end{align}
The estimator in Eq.~\eqref{eq:fnl_estimator} then becomes
\begin{align}
	\hat f_{\rm NL} = \frac{(2\pi)^3}{N_{\rm th}} \int \frac{\d^3k}{(2\pi)^3}   \frac{h(-\k)\delta(-\k)}{P_\delta(k)}\int^\prime_\q   \frac{f(\q)\delta(\q)}{P_\delta(q)}\frac{g(\k-\q)\delta(\k-\q)}{P_\delta(|\k-\q|)}\, , \label{fNLestimator}
\end{align}
 where the notation of $\int^\prime _\q$ refers to the 3-dimensional integration $\int \d^3 q/(2\pi)^3$, imposing the condition that only modes with $q , |\k-\q|< k_{\rm max}$ contribute. This condition is set to avoid contributions from  small-scales modes to the skew-spectrum, and to make the analysis consistent with the standard analysis of the bispectrum where the condition of $k_i < k_{\rm max}$ is imposed for all wavenumbers $k_i$.\footnote{Alternatively, other types of smoothing can be used to suppress small-scale modes, e.g.~a Gaussian smoothing~\cite{Schmittfull:2014tca}, but we use a sharp filter here to ensure that we use exactly the same modes as the standard bispectrum analysis.}
As is expected, choosing a larger value of $k_{\rm max}$ generally increases the signal-to-noise, with the trade-off that we approach a scale at which our theoretical model becomes no longer accurate. Now, writing the integral over $\q$ as a convolution of the filtered densities
\begin{align}\label{eq:conv}
	\bigg[\frac{f\delta}{P_\delta}\star\frac{g\delta}{P_\delta}\bigg](\k) \equiv \int^\prime_\q  \frac{f(\q)\delta(\q)}{P_\delta(q)}\frac{g(\k-\q)\delta(\k-\q)}{P_\delta(|\k-\q|)}\, ,
\end{align}
we can express the estimator in the following compact form \cite{Schmittfull:2014tca}: 
\begin{align}
	\hat f_{\rm NL} 
	&=\frac{4\pi V_i}{N_{\rm th}}\int \frac{k^2\d k}{P_\delta(k)}\,\hat P_{\frac{f\delta}{P_\delta}\star\frac{g\delta}{P_\delta}, h\delta}(k)\, ,\label{fNLestimator2}
\end{align}
where
\begin{eBox}
\vskip 4pt
\be\label{eq:cross}
\hat P_{\frac{f\delta}{P_\delta}\star\frac{g\delta}{P_\delta}, h\delta}(k) \equiv \frac{1}{V_i V_s(k)} \int_k \d^3k_1 \left[\frac{f\delta}{P_\delta}\star\frac{g\delta}{P_\delta}\right] (\k_1) \left[h \delta\right](-\k_1)\, ,
\ee
\vskip 3pt
\end{eBox}
defines the skew-spectrum estimator between the quadratic field~\eqref{eq:conv} and the filtered density field $[h\delta](\k)\equiv h(k)\delta(\k)$. The integral $\int_k$ is over modes within the shell centered at $k$ with the volume $V_s(k) = 4 \pi k^2 \Delta k$, and $V_i$ is the volume of the redshift bin in the range $ z \in[z_{\rm min},z_{\rm max}]$, which for a survey with a sky coverage of $f_{\rm sky}$ is given in terms of comoving distance $d_c$ as 
\be
V_i = \frac{4\pi f_{\rm sky}}{3} \big[ d_c^3(z_{\rm max}) - d_c^3(z_{\rm min})\big]\, .
\ee 

\noindent For the skew-spectra corresponding to the bispectrum due to gravitational evolution, the $k^2/P_\delta(k)$ weighting of the optimal bispectrum estimator in Eq.~\eqref{fNLestimator2} corresponds to the inverse variance weighting of the cross-spectrum on large scales, assuming Gaussian bispectrum covariance. This is due to the fact that in this case the variance of the cross-spectra on large scales, $k\rightarrow0$, scales like $P_\delta(k)/k^2$ \cite{Schmittfull:2014tca}. 
 
If the theoretical bispectrum has a separable form, the maximum likelihood estimate of its amplitude can therefore be written as the skew-spectrum estimator of appropriately filtered quadratic and linear density fields. In practice, the convolution of the two filtered fields in Eq.~\eqref{eq:conv} can be efficiently performed by first multiplying the density in Fourier space with the two filters $f/P$ and $g/P$, then inverse-Fourier transforming the filtered fields to configuration space and multiplying them, and finally Fourier transforming the result back to Fourier space. The computational cost is dominated by the cost of Fourier transforms, which if evaluated using Fast Fourier transforms (FFTs) requires $\O(N\log N)$ operations, where $N=(k_{\rm max}/k_{\rm min})^3$ is the number of  3D grid points at which the fields are evaluated. For example, $N=10^6$ for 100 grids points per dimension. This means that the full information of the bispectrum on its amplitude can be computed in $\O(N\log N)$ time using the skew-spectrum defined by Eq.~\eqref{eq:cross}.
The computational cost is thus equivalent to calculating the power spectrum, and avoids the evaluation of all triangular configurations which requires $\O(N^2)$ operations. 

If the bispectrum model is a sum of separable terms, we can compute one skew-spectrum for each such term to determine the amplitude of each contribution. 
By combining these skew-spectra, we obtain the full bispectrum information on such amplitude-like parameters, such as galaxy biases and the nonlinear parameter $f_{\rm NL}$ for primordial non-Gaussianity. Constructing these skew-spectra will be the subject of the subsequent sections.

\section{Skew-Spectra for the Tree-Level Bispectrum}\label{sec:theory} 
In this section, we construct the skew-spectra corresponding to the tree-level matter and galaxy bispectrum, accounting for the contributions from primordial non-Gaussianity (\S\ref{sec:cross_png}) and gravitational evolution (\S\ref{sec:cross_grav}). In \S\ref{sec:cross_shot}, we compute the shot-noise contribution to each of the skew-spectra, arising from the stochastic contribution to the galaxy bispectrum. Lastly, we compute the covariances of the skew-spectra which are needed for performing Fisher analysis and obtaining parameter constraints in \S\ref{sec:covs}.

\subsection{Primordial Non-Gaussianity}\label{sec:cross_png}

First, we construct the skew-spectra arising from non-Gaussian initial conditions. As a case study, we outline a detailed construction of the skew-spectrum corresponding to the primordial bispectrum due to couplings of the inflaton field to extra particles with generic spins and masses comparable to the Hubble scale, in the context of ``cosmological collider physics''~\cite{Chen:2009zp, Baumann:2011nk, Assassi:2012zq, Noumi:2012vr, Assassi:2013gxa, Arkani-Hamed:2015bza, Chisari:2016xki, Lee:2016vti, Kehagias:2017cym, Kumar:2017ecc, An:2017hlx, An:2017rwo, Baumann:2017jvh, Bordin:2018pca, Arkani-Hamed:2018kmz, Baumann:2019oyu, Wang:2019gbi}, along with other conventional shapes. After discussing the separability of the primordial bispectrum in \S\ref{sec:septemp}, we derive the corresponding cross-spectra using the maximum-likelihood $f_{\rm NL}$ estimator for the matter and galaxy bispectra in \S\ref{sec:skewmatter} and \S\ref{sec:skewgal}, respectively. 

\subsubsection{Separable Templates}\label{sec:septemp}
The bispectrum of the primordial curvature perturbation $\zeta$ is defined as \be
\langle \zeta(\k_1)\zeta(\k_2)\zeta(\k_3)\rangle = (2\pi)^3 \delta_{\rm D}(\k_1+\k_2+\k_3) B_\zeta(k_1,k_2,k_3)\, .
\ee
To better illustrate how the construction of the skew-spectrum \eqref{eq:cross} works in practice, we will study several specific primordial bispectra in our analysis. For example, the local and equilateral bispectra are given by \cite{Gangui:1993tt,Wang:1999vf,Verde:1999ij,Komatsu:2001rj,Babich:2004gb,Creminelli:2005hu}
\begin{align}
	B_\zeta^{\rm loc}(k_1,k_2,k_3) &= \frac{6}{5} f_{\rm NL}^{\rm loc}\Big[ P_\zeta(k_1)P_\zeta(k_2) + 2\ {\rm perms}\Big]\, ,\label{Bloc} \\[3pt]
	B_\zeta^{\rm eq}(k_1,k_2,k_3) &= \frac{18}{5} f_{\rm NL}^{\rm eq} \Big[{-} 2P_\zeta^{2/3}(k_1)P_\zeta^{2/3}(k_2)P_\zeta^{2/3}(k_3) -\Big( P_\zeta(k_1) P_\zeta(k_2) + 2\ {\rm perms}\Big)\nonumber \\[3pt]
	&\hspace{.6in} + \Big(P_\zeta(k_1)^{1/3}P_\zeta(k_2)^{2/3}P_\zeta(k_3) + 5 \ {\rm perms} \Big)\Big]\,.\label{eq:eqtemp}
	\end{align}
The local non-Gaussianity typically arises in multi-field inflationary scenarios, and \eqref{Bloc} corresponds to having a local expansion of Gaussian fields in real space. 
The equilateral template captures the shape that arises from the interactions $\dot\pi^3$ and $\dot\pi(\partial_i\pi)^2$, where $\pi$ is the Goldstone boson of broken time translations during inflation~\cite{Cheung:2007st}. These describe the self-interactions of the scalar fluctuations during inflation at leading order in derivatives. If these fluctuations couple to other particles present during inflation, then other distinctive shapes will be generated that depend on the masses and spins of the extra particles. Although these shapes are difficult to compute for general momentum configurations, they take simple functional behaviors in certain limits. For instance, they exhibit an oscillatory behavior with angular dependence in the squeezed limit, whereas away from it the shape degenerates with the higher-spin version of the equilateral shape~\cite{Arkani-Hamed:2015bza, Lee:2016vti, Arkani-Hamed:2018kmz}.\footnote{Physically, this is because the non-local effect due to massive particles only shows up when the distance between two correlated points becomes largely separated, which translates in momentum space to the soft limit of the momentum conjugate to this distance. Around the equilateral configuration, the shape is well-approximated by an interaction that arises from integrating out the massive particle, e.g.~$\dot\pi(\partial_{i_1\cdots i_s}\pi)^2-\text{traces}$ for a spin-$s$ particle, which still retains the spin dependence.} Using this information, Ref.~\cite{MoradinezhadDizgah:2018ssw} introduced a template for the primordial bispectrum arising from massive particles with spin, denoted as $B_\zeta^{(s)}$, which takes the form
\be\label{eq:Bfull}
B^{(s)}_\zeta(k_1,k_2,k_3) =  \alpha^{(s)}\hskip -1pt f_{\rm NL}^{(s)}\big[B^{\small\rm A}(k_1,k_2,k_3)+ B^{\text{NA}}_\nu(k_1,k_2,k_3)\big]\, ,
\ee
where $B^{\rm A}$ and $B^{\rm NA}$ respectively capture the information about the ``EFT expansion'' and ``particle production'' due to the particle exchange process. These are called the ``analytic'' and ``non-analytic'' parts that reflect their scaling behavior in the squeezed limit. For detecting extra particles, we are interested in measuring the latter contribution. A spin-dependent prefactor $\alpha{(s)}$ is inserted to ensure the standard normalization in the equilateral configuration, $f_{\rm NL}^{(s)} =\frac{5}{18}B^A(k,k,k)/P_\zeta^2(k)$. The precise shapes are given by 
\begin{align}
B^{\rm A}(k_1,k_2,k_3)  &= \frac{\hskip -0.5pt\L_s(\hat \k_1\cdot\hat \k_2)(k_1k_2)^{s}}{k_3 k_t^{2s+1}}\Big[(2s-1)\big((k_1+k_2)k_t+2sk_1k_2\big)+k_t^2 \Big] P_\zeta(k_1)P_\zeta(k_2) \nn
&+ \text{2 perms}\, ,\label{Bangle}\\[3pt]
B^{\text{NA}}_\nu(k_1,k_2,k_3)  &= r^{(s)}\hskip -0.5pt(\nu)\left(\frac{k_1}{k_2}\right)^{3/2}\cos\left[\nu\ln\frac{k_1}{k_2}+\varphi\right]\! \L_s(\hat\k_1\cdot\hat\k_2)  \Theta(x_*k_2-k_1)P_\zeta(k_1)P_\zeta(k_2) \nn
&+ \text{5 perms}\, ,\label{BNA}
\end{align}
where $k_t \equiv k_1+k_2+k_3$ is the total wavenumber, $\nu \equiv \sqrt{(m/H)^2-(s-1/2-\delta_{s0})^2}$ is the mass parameter with $m$ being the mass of the particle, $\L_s$ is the Legendre polynomial of degree $s$, and $\Theta$ is the Heaviside step function, which for a squeezing factor $x_*<1$ accounts for the fact that the non-analytic template is only valid in the squeezed-limit. In practice, we set $x_*=0.1$. For a given type of interaction, the phase $\varphi$ is fully determined in terms of the mass parameter $\nu$; for simplicity, we will set $\varphi=0$. An important parameter is $r^{(s)} (\nu)$, which sets the relative size of the analytic and non-analytic parts for a given spin. For large $\nu$, it goes as $r^{(s)}\sim e^{-\pi\nu}$ due to the Boltzmann suppression for pair-production of particles in a near-de Sitter inflationary background. This implies that the signal will be dominated by the analytic part, unless $\nu$ is not too bigger than one. For the explicit expressions of $r^{(s)}$ up to $s=4$, see~\cite{MoradinezhadDizgah:2018ssw}.

The local and equilateral templates are manifestly in a separable form. In contrast, the dependence on $k_t$ in the denominator of Eq.~\eqref{Bangle} and the cosine and Heaviside functions in Eq.~\eqref{BNA} make the template for particle exchange in Eq. \eqref{eq:Bfull} naively look non-separable. However, there is a well-known simple trick to make the $k_t$ part separable, which is by the use of the Schwinger parameterization~\cite{Smith:2006ud, Oppizzi:2017nfy}
 \be\label{eq:Schwinger}
 \frac{1}{k_t^n} = \frac{1}{\Gamma(n)} \int_0^\infty \d\tau \, \tau^{n-1} e^{-\tau k_t}\, .
 \ee
 In rewriting the analytic part of the template, since the integrand is a smooth function, the integration over $\tau$ can be well-approximated as a sum, requiring typically of order $\O(10-10^2)$ terms to reach a percent-level accuracy. Writing the argument of the Legendre polynomial as $\hat \k_1\cdot\hat \k_3=(k_2^2-k_1^2-k_3^2)/2$, we see that the analytic part of the template can be written in a manifestly separable form.

For the non-analytic part of the template, we note that the cosine in Eq.~\eqref{BNA} can be written as a sum over two complex power-laws
\be
2\cos\left[\nu\ln\left(\frac{k_1}{k_3}\right) \right] = \left( \frac{k_1}{k_3}\right)^{i \nu}  +\left( \frac{k_1}{k_3}\right)^{-i \nu}\, ,
\ee
which are manifestly separable. Moreover, the naively non-separable Heaviside function can be dealt with by approximating it with the following sum~\cite{Dominici_2012, Slepian:2018vds}
\begin{align}
	\Theta(x_*k_3-k_1) \approx  \frac{2(x_*k_3)^2}{\pi}\sum_{n=1}^{N_\Theta} n j_0(n x_*k_3/k_\star)j_1(nk_1/k_\star)\, ,\label{thetasum}
\end{align}
where the sum converges for $0<x_*k_3-k_1<2\pi k_\star$ and we have introduced an arbitrary scale $k_\star$ to make up the dimensions. This sum converges very fast for $x_*k_3\ll k_1$ and requires $N_\Theta=\O(10^2)$ terms near the cutoff, $x_*k_3\approx k_1$, for a percent-level precision. 

Instead of following the above procedure to rewrite the analytic part of the template in an explicitly separable form, we can also construct approximate separable templates. For example, the equilateral template \eqref{eq:eqtemp} was constructed precisely in this manner. We discuss the construction of such a template and its accuracy compared to the full template in Appendix~\ref{sec:sep_temp}. Having written the bispectrum templates in separable forms, next we derive the nonlinear kernels and the skew-spectra of the matter and halo bispectra following from the above three primordial bispectra. 

\subsubsection{Skew-Spectra for the Matter Bispectrum}\label{sec:skewmatter}
The matter bispectrum due to the presence of primordial non-Gaussianity with nonzero bispectrum $B_\zeta$ is given by
\be
B^{\rm PNG}_m(k_1,k_2,k_3,z) =  \ {\mathcal M}(k_1,z){\mathcal M}(k_2,z) {\mathcal M}(k_3,z)  B_\zeta(k_1,k_2,k_3)\, ,
\ee
where ${\cal M}(k,z)$ is the transfer function that relates the primordial fluctuations $\zeta$ to the linearly extrapolated matter overdensity $\delta_0$ during the matter-domination era 
\begin{align}
	\delta_0(\k,z) = {\mathcal M}(k,z)\zeta(\k)\, , \quad
{\mathcal M}(k,z) = -\frac{2}{5} \frac{k^2 T(k)D(z)}{\Omega_m H_0^2}\, ,
\end{align}
with $T(k\rightarrow 0) =1$. To avoid clutter, from here on we drop the explicit redshift dependence of the transfer function.

For a general separable primordial bispectrum, the estimator for the skew-spectrum, averaged over the Fourier bin of width $\Delta k$ centered at $k$, is given in terms of the cross-spectra of the quadratic and linear filtered matter fields $\delta_m(\k)/{\mathcal M}(k)$ by~\cite{Schmittfull:2014tca}
\be \label{eq:estimator}
\hat P_{D^{\rm PNG}\left[\frac{\delta_m}{\mathcal M}\right],\frac{\delta_m}{\mathcal M}}(k) = \frac{1}{V_i V_s(k)} \int_k \d^3k_1  \ D^{\rm PNG}\left[\frac{\delta_m}{\mathcal M}\right] (\k_1)\left[\frac{\delta_m}{\mathcal M}\right](-\k_1)\, ,
\ee
where we defined the quadratic filtered field corresponding to the primordial bispectrum as 
\be\label{eq:mcross_png}
D^{\rm PNG}\left[\frac{\delta_m}{{\mathcal M}} \right](\k) = \int^\prime_\q  D^{\rm PNG}(\q,\k-\q) \frac{P_0(q)P_0(|\k-\q|)}{P_m(q)P_m(|\k-\q|)} \frac{\delta_m(\q)}{{\mathcal M}(q)} \frac{\delta_m(\k-\q)}{{\mathcal M}(|\k-\q|)}\,,
\ee
with $P_0$ being the linear matter power spectrum, and again we have dropped the explicit redshift dependence of $P_0$. The kernel $D^{\rm PNG}$ is the non-linear kernel determined by the shape of primordial bispectrum under consideration. Below we will derive the form of this kernel for the local, equilateral, and the spin-$s$ exchange bispectrum templates, which we will denote by $D^{\rm loc}$, $D^{\rm eq}$, and $D^{\rm spin}$, respectively. 

In the above definition of the quadratic field, the ratio of the linear matter power spectrum to that of the observed density field, $P_0/P_m$, plays the role of a weighting function.
For dark matter, the weight becomes unity on large scales. Setting this ratio to unity on all scales amounts to choosing a sub-optimal weight on small scales due to the nonlinear corrections to the power spectrum and shot noise for discrete tracers. It was argued in Ref.~\cite{Schmittfull:2014tca} that the drop in the integrand due to the smoothing kernel used in calculating the cross-spectra is much faster than the drop of the ratio of the power spectra, and the weight was therefore set to unity. However, we will instead keep the optimal weight in our analysis. When considering the cross-spectra corresponding to the halo bispectrum, the power spectra in the denominator are replaced by the power spectrum of halos, including shot noise.

Taking the expectation value of the estimator gives the skew-spectrum\footnote{Since we neglect redshift space distortions and assume that the primordial bispectrum is isotropic, the matter and  galaxy bispectrum are a function only of $k_1$, $k_2$ and $k_3$. The skew-spectrum is therefore a function only of $k$.}
\be
P_{D^{\rm PNG}\left[\frac{\delta_m}{\mathcal M}\right],\frac{\delta_m}{\mathcal M}}(k)  = \int^\prime_\q \frac{P_0(q)P_0(|\k-\q|)}{P_m(q)P_m(|\k-\q|)} \frac{D^{\rm PNG}(\q,\k-\q)  
B_m(q,|\k-\q|,k)
}{{{\mathcal M}(q) {\mathcal M}(|\k-\q|)\mathcal M}(k)}\, ,
\ee
where $B_m$ is the matter bispectrum. At tree-level in perturbation theory, this is 
\be
B_m(k_1,k_2,k_3) =  B_m^{\rm PNG} (k_1,k_2,k_3)+  B_m^{\rm grav}(k_1,k_2,k_3)\, ,
\ee
where the tree-level matter bispectrum due to gravitational evolution is 
\be\label{eq:mbis}
B_m^{\rm grav}(k_1,k_2,k_3) = 2 F_2(\k_1,\k_2) P_0(k_1) P_0(k_2) +2\ {\rm perms}\, ,
\ee
and $F_2$ is the matter mode-coupling kernel~\cite{Goroff:1986ep, Bernardeau:2001qr}
\be
F_2(\k_1,\k_2) \equiv \frac{5}{7} + \frac{\k_1\cdot\k_2}{2k_1k_2}\left(\frac{k_1}{k_2} + \frac{k_2}{k_1} \right) + \frac{2}{7}\left(\frac{\k_1\cdot\k_2}{k_1k_2}\right)^2.
\ee
Note that we have made the assumption that upon taking the ensemble average in Eq.~\eqref{eq:estimator}, the integrand is constant within each $k$-bin for narrow bin widths.

The nonlinear kernel for the primordial bispectrum in Eq.~\eqref{eq:Bfull} can be written as 
\be
D^{\rm spin}(\q,\k-\q) = 3\sum_i^3 D_i^{(s)}(\q,\k-\q) + 6 D^{(\nu)} (\q,\k-\q)\, ,
\ee
where the kernels $D_i^{(s)}$ correspond to the analytic part of the template, while the kernel $D^{(\nu)}$ corresponds to the non-analytic part and are given by 
\begin{align}\label{eq:a_kernels}
D_1^{(s)}(\q,\k-\q) &= \frac{(2s-1) \left(|\k-\q|^s q^{1+s} +|\k-\q|^{1+s} q^s \right)}{k\left(k+q+|\k-\q|\right)^{2s}}  \L_s(\mu)\, , \nonumber \\
D_2^{(s)}(\q,\k-\q) &= \frac{2s(2s-1)|\k-\q|^{1+s} q^{1+s}}{k\left(k+q+|\k-\q|\right)^{2s+1}}  \L_s(\mu)\, , \nonumber \\
D_3^{(s)}(\q,\k-\q) &= \frac{|\k-\q|^{s} q^{s}}{k\left(k+q+|\k-\q|\right)^{2s-1}} \L_s(\mu)\, .  
\end{align} 
For the non-analytic part, we have 
\begin{align}\label{eq:na_kernels}
D^{(\nu)}(\q,\k-\q) =  r^{(s)}\hskip -0.5pt(\nu) \left(\frac{|\k-\q|}{q}\right)^{3/2} \cos\left[\nu\ln\left(\frac{|\k-\q|}{q}\right)\right]\Theta(x_*q-|\k-\q|)\L_s(\mu)\, .
\end{align} 
Following a similar derivation, the nonlinear kernels corresponding to the primordial bispectrum of the local and equilateral shape are  \begin{align}
	D^{\rm loc} (\q,\k-\q) & = 1\, , \\
	D^{\rm eq} (\q,\k-\q) & =  -3 - \frac{2 P_\zeta^{2/3}(k)}{P_\zeta^{1/3}(q)P_\zeta^{1/3}(|\k-\q|)} +   \frac{6 P_\zeta(k)}{P_\zeta^{1/3}(q) P_\zeta^{2/3}(|\k-\q|)}\, ,
\end{align}
\vspace{.03in}
where we used the symmetry $\q\leftrightarrow\k-\q$ in the integrand to combine different terms.

\subsubsection{Skew-Spectra for the Galaxy Bispectrum}\label{sec:skewgal}
We use a simple bias prescription in Eulerian space, where the galaxy overdensity at point $\x$ is expanded in terms of the matter overdensity and the traceless part of the tidal tensor at the same location. Up to quadratic order, we have
\be
\delta_g = b_1 \delta_m + b_2 \delta_m^2 +  b_{K^2}[K_{ij}]^2, \label{eq:gbias}
\ee
where $K_{ij}$ is the tidal tensor defined as 
\be
K_{ij}({\bf x}) \equiv \left( \frac{\partial_i \partial_j}{\partial^2} - \frac{1}{3} \delta_{ij}\right) \delta_m({\bf x})\, .
\ee
For the smoothed galaxy density field, the estimator for the skew-spectrum is defined similarly to Eq.~\eqref{eq:mcross_png}, but with the matter density field $\delta_m$ replaced by $\delta_g$. The expectation value of the estimator is
\be\label{eq:gcross_PNG}
P_{D^{\rm PNG}\left[\frac{\delta_g}{\mathcal M}\right],\frac{\delta_g}{\mathcal M}}(k)  = \int^\prime_\q \frac{P_0(q)P_0(|\k-\q|)}{P_g(q)P_g(|\k-\q|)} \frac{D^{\rm PNG}(\q,\k-\q)  B_g(k,q,|\k-\q|)}{{\mathcal M}(q) {\mathcal M}(|\k-\q|){\mathcal M}(k)}\, .
\ee
Accounting for the contributions from gravitational evolution and primordial bispectrum, the galaxy bispectrum at tree-level in perturbation theory is given by
\be\label{eq:bis_tot}
B_g(k_1,k_2,k_3) =  B^{\rm PNG}_g(k_1,k_2,k_3) + B^{\rm grav}_g(k_1,k_2,k_3)\, ,
\ee
where
\begin{align}
B^{\rm grav}_{\rm g}(k_1,k_2,k_3) &= 2 \big[  b_1^3 F_2(\k_1,\k_2) +b_1^2 b_2 + b_1^2 b_{K^2} K_2(\k_1,\k_2) \big]  P_0(k_1) P_0(k_2) + 2 \ {\rm perms}\, \label{eq:gbis}, \\[3pt]
B^{\rm PNG}_{\rm g}k_1,k_2,k_3) &=  b_1^3 \ {\mathcal M}(k_1){\mathcal M}(k_2) {\mathcal M}(k_3)  B_\zeta(k_1,k_2,k_3)\, \label{eq:gbis_png},
\end{align}
with $K_2$ being the square of the tidal field in Fourier space
\be
K_2(\k_1,\k_2) \equiv \left(\frac{\k_1\cdot\k_2}{k_1k_2}\right)^2 -\frac{1}{3}\, .
\ee
To make the notation more concise, in the rest of the paper we will redefine the nonlinear kernel corresponding to the primordial bispectrum of matter or galaxies as
\be\label{eq:NL_PNG}
\tilde D^{\rm PNG}(\q,\k-\q) \equiv \frac{P_0(q)P_0(|\k-\q|)}{P_a(q)P_a(|\k-\q|)} \frac{D^{\rm PNG}(\q,\k-\q)}{{\mathcal M}(k){\mathcal M}(q) {\mathcal M}(|\k-\q|)}\, ,
\ee
and use the notation
\be
P_{\tilde D^{\rm PNG}[\delta_a],\delta_a}(k)  = P_{D^{\rm PNG}\left[\frac{\delta_a}{\mathcal M}\right],\frac{\delta_a}{\mathcal M}}(k) \, ,
\ee
with the subscript $a\in\{m,g\}$ denoting either the matter or galaxy density field. We will show the shape of the skew-spectrum due to local non-Gaussianity, as well as those due to gravitational evolution in Fig.~\ref{fig:skew_spec}. 

We note that a nonzero primordial bispectrum can induce additional contributions the bispectrum of biased tracers beside Eq.~\eqref{eq:gbis_png}. These contributions are due to the fact that in the presence of primordial non-Gaussianity, the bias expansion in Eq.~\eqref{eq:gbias} is modified and new bias operators are needed~\cite{Assassi:2015fma,Desjacques:2016bnm}. For instance, at linear level, the observed density field of the biased tracer would depend on the gravitational potential at early times $\phi$ in addition to matter density field. Therefore, to capture the full information of the bispectrum, in addition to the non-linear kernel in Eq.~\eqref{eq:gbis_png}, one need to construct new kernels to capture the shape of these new contribution. 

As we discuss in Appendix~\ref{app:full_bis_loc}, depending on the shape of the primordial bispectrum, accounting for the additional contributions may be important. In particular, for the local shape, the linear dependence on $\phi$ gives rise to the so-called {\it scale-dependent bias} in the galaxy power spectrum~\cite{Dalal:2007cu, Matarrese:2008nc,Afshordi:2008ru}, as well as in the bispectrum. 
In this work, we neglect these contributions here and only consider the non-linear kernels corresponding to Eq.~\eqref{eq:gbis_png}. While this is a good assumption for the equilateral and spin-exchange bispectra, the new contributions should be correctly accounted for local non-Gaussianity. We defer this to a future analysis.

\subsection{Gaussian Initial Conditions}\label{sec:cross_grav}
The skew-spectra corresponding to the matter and halo bispectra due to gravitational evolution of Gaussian initial conditions can be derived similarly to the ones above for primordial non-Gaussianity. 
This leads to three types of skew-spectra between a quadratic and a linear density field \cite{Schmittfull:2014tca}. The three quadratic fields in configuration space are the squared density field $\delta^2(\x)$, a shift term that contracts the displacement and gradient of the density $\psi^i(\x)\partial_i\delta(\x)$, and a tidal shear term $[K_{ij}(\x)]^2$. In this section, we simply summarize the final expressions from Ref.~\cite{Schmittfull:2014tca} for the expectation values of the matter and halo cross-spectra that we will use in our Fisher forecast, and refer the interested reader to Ref.~\cite{Schmittfull:2014tca} for more details. 

\subsubsection{Skew-Spectra for the Matter Bispectrum}
The skew-spectra corresponding to the maximum likelihood estimator of the amplitude of the matter bispectrum from Gaussian initial conditions, given by Eq.~\eqref{eq:mbis}, takes the form~\cite{Schmittfull:2014tca}
\begin{align}\label{eq:mcross_g}
P_{D^{\rm grav}[\delta_m],\delta_m }(k) &= \int^\prime_\q \frac{P_0(q)P_0(|\k-\q|)}{P_m(q)P_m(|\k-\q|)} D^{\rm grav}(\q,\k-\q) B_m(\q,\k-\q,-\k) \nonumber \\[3pt]
&= 2 I_{\tilde D^{\rm grav}F_2}(k) + 4 I_{\tilde D^{\rm grav}F_2}^{{\rm bare}}(k)\, ,
\end{align}
where the nonlinear kernels $D^{\rm grav}$ corresponding to the three quadratic fields $\delta_m^2, \Psi^i_m\partial_i\delta_m$ and $\left[K_{ij}\right]_m^2$ are $D^{\rm grav}(\k_1,\k_2) \in \{\L_0(\hat \k_1\cdot\hat \k_2), F_2^1 (k_1,k_2)\L_1(\hat \k_1\cdot\hat \k_2), \L_2(\hat \k_1\cdot\hat \k_2)\}$, with the symmetric kernel $F_2^1$ defined as 
\be
F_2^1(k_1,k_2) = \frac{1}{2}\left(\frac{k_1}{k_2} + \frac{k_2}{k_1}\right).
\ee
The three non-linear kernels $D^{\rm grav}$ correspond to the matter bispectrum, decomposing the symmetrized kernel for the second-order density perturbations, $F_2$, in terms of the Legendre polynomials. The tilde in the second line of Eq.~\eqref{eq:mcross_g} denotes absorption of the ratio of the power spectra in the nonlinear kernel, similar to the case of nonlinear kernels corresponding to primordial bispectrum in Eq.~\eqref{eq:NL_PNG}
\be \label{eq:NL_G}
\tilde D^{\rm grav}(\q,\k-\q) =  \frac{P_0(q)P_0(|\k-\q|)}{P_a(q)P_a(|\k-\q|)} D^{\rm grav}(\q,\k-\q)\,,
\ee
with the subscript $a\in\{m,g\}$ again referring to either matter or galaxy density field. The two $I$ functions are given by
\begin{align}\label{eq:Ifuncs}
I_{\tilde D^{\rm grav}F_2}(k) &\equiv  \int^\prime_\q \tilde D^{\rm grav}(\q,\k-\q) F_2(\q,\k-\q) P_0(q) P_0(\k-\q|)\, , \nonumber \\
 I_{\tilde D^{\rm grav}F_2}^{{\rm bare}}(k) &\equiv  \int^\prime_\q  \tilde D^{\rm grav}(\q,\k-\q)F_2(\q,-\k)P_0(q)P_0(k)\,.
\end{align}
Note that $I_{{\tilde D}F}$ is symmetric under the exchange $\tilde{D} \leftrightarrow F$, while $I_{ \tilde{D}F}^{\rm bare}$ is not.

\subsubsection{Skew-Spectra for the Galaxy Bispectrum}
Similarly, when considering the galaxy bispectrum  due to the gravitational evolution of Gaussian initial conditions in Eq.~\eqref{eq:bis_tot}, the expectation value of the maximum likelihood estimator of the bias parameters (amplitudes) can be written in terms of three skew-spectra of the form \cite{Schmittfull:2014tca}
\bea\label{eq:gcross}
 P_{ \tilde D^{\rm grav} [\delta_g], \delta_g}(k) &=& 2b_1^3\left[ I_{\tilde D^{\rm grav}F_2}(k) + 2 I_{\tilde D^{\rm grav}F_2}^{\rm bare} (k) \right] 
+ 2b_1^2b_2\left[I_{\tilde D^{\rm grav}\L_0}(k) + 2 I_{\tilde D^{\rm grav}\L_0}^{\rm bare} (k) \right] \nonumber \\
&+& \frac{4}{3}b_1^2b_{K^2} \left[I_{\tilde D^{\rm grav}\L_2}(k) + 2 I_{\tilde D^{\rm grav}\L_2}^{\rm bare} (k) \right],
\eea
where the quadratic kernels are again given by $D^{\rm grav}(\k_1,\k_2) \in \{\L_0(\hat \k_1\cdot\hat \k_2)$, $F_2^1 (k_1,k_2)\L_1(\hat \k_1\cdot\hat \k_2)$,  $\L_2(\hat \k_1\cdot\hat \k_2)\}$. The functions $ I_{\tilde D^{\rm grav}\L_i}$ and $ I_{\tilde D^{\rm grav}\L_i}^{\rm bare}$ are defined as in Eq.~\eqref{eq:Ifuncs}, after replacing the $F_2$ kernel with the Legendre polynomials $\L_0$ and $\L_2$.

\begin{figure}[t]
\centering
\hspace{-.6in}\includegraphics[width = .7\textwidth]{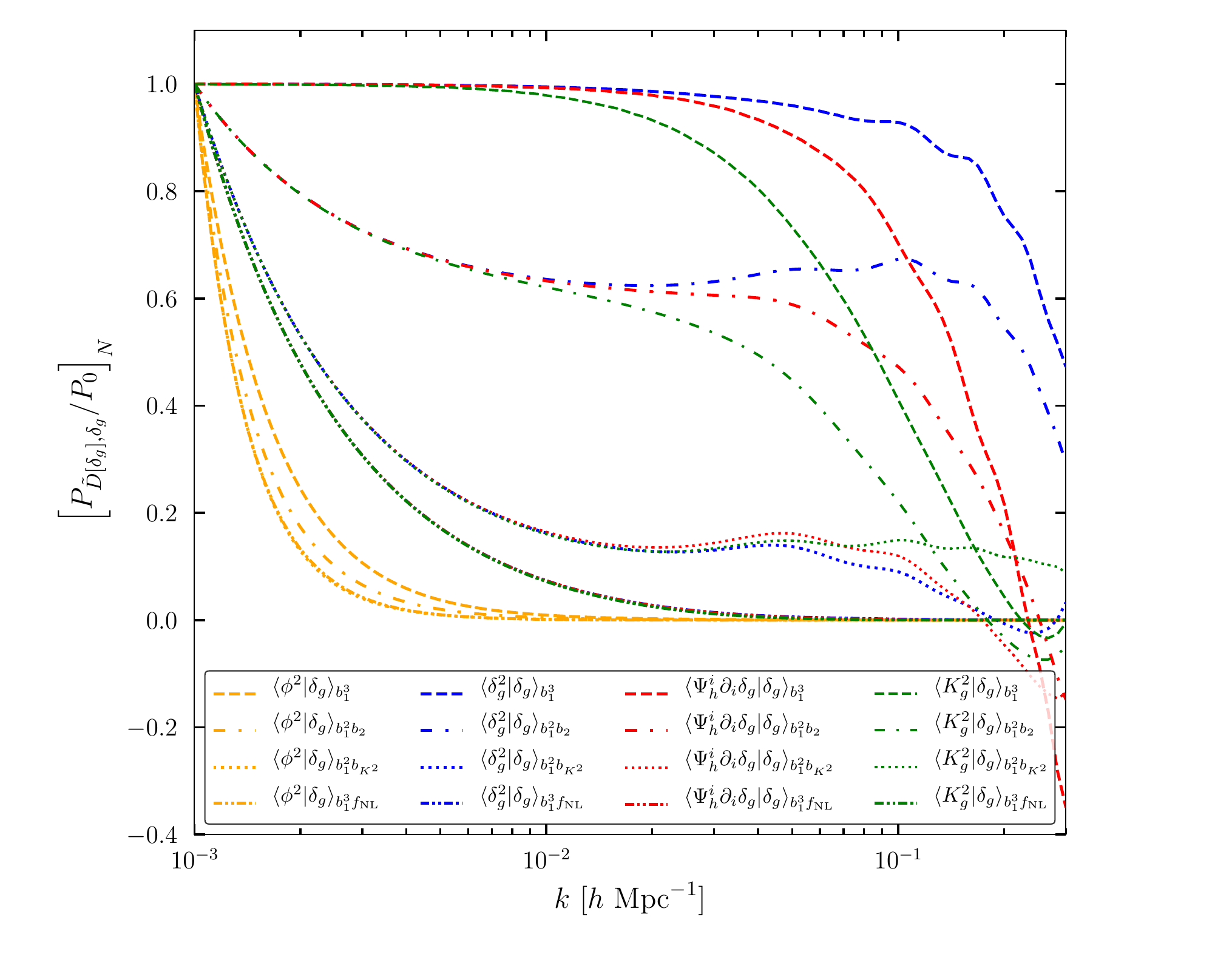}
\caption{The shapes of various contributions to the skew-spectra corresponding to the tree-level galaxy bispectrum given in Eqs.~(\ref{eq:bis_tot}-\ref{eq:gbis_png}) at $z=1$. For plotting convenience, we divide the skew-spectra by the linear matter power spectrum $P_0$. Furthermore, since the overall amplitudes of the skew-spectra due to primordial and gravitational bispectra are significantly different, to show the shapes of different contributions at large and small scales more clearly, we normalized $P_{\tilde D[\delta_h],\delta_h}/P_0$ to be unity at $k = 0.001 \ h \ {\rm Mpc}^{-1}$ (hence the subscript $N$ on the y-axis). We have grouped the contributions based on their scaling with biases and $f_{\rm NL}$; scaling as $b_1^3$ (dashed), $b_1^2 b_2$ (dashed-dotted), $b_1^2 b_{K^2}^2$ (dotted) and $b_1^3 f_{\rm NL}$ (dashed-double-dotted). The orange line  correspond to the quadratic kernel $\tilde D^{\rm PNG}$ for the local shape, and the field $\phi$ denotes the gravitational potential at initial conditions. The blue, red and green lines correspond to the skew spectra constructed from the three quadratic kernels for $\tilde D^{\rm grav}$, corresponding to squared density $\delta_g^2$, shift term $\Psi^i_h\partial_i \delta_g$ and tidal term $K_g^2$. In computing the weight (the ratio of the linear matter power spectra to the observed power spectra), we used the DESI shot-noise corresponding to redshift bin centered at $z=1$.}
\label{fig:skew_spec}
\end{figure}

In Fig.~\ref{fig:skew_spec}, we show the shapes of the contributions to the skew-spectra characterizing the tree-level galaxy bispectrum in the presence of local non-Gaussianity. Similar to Ref.~\cite{Schmittfull:2014tca}, we have grouped the contributions according to their scaling with bias parameters and $f_{\rm NL}$. As we discussed earlier, to discard the contribution from small-scale modes, the skew-spectra are computed using a sharp top-hat filter in Fourier space. Different lines and colors are described in the caption of the figure. As was shown in Ref.~\cite{Schmittfull:2014tca} for the skew-spectra due to gravitational evolution, different terms with the same scaling with biases have the same shape at large scales. When accounting for local non-Gaussianity, all contributions asymptotically approach zero at small scales. On large scales, the contributions that scale as $b_1^3 f_{\rm NL}$ and are sourced by non-primordial quadratic fields (blue, red and green lines) differ negligibly in their shapes, while the term sourced by the quadratic primordial field (yellow dashed-double-dotted line) has a different shape. The shape of the contributions from the primordial quadratic field scaling with other bias combinations (dashed, dashed dotted, and dotted yellow lines) also differ in their shapes. The two terms scaling as $b_1^2 b_{K^2}$ and $b_1^3 f_{\rm NL}$ (yellow dotted and dashed-dotted lines) nearly overlap.

\subsection{Shot Noise}\label{sec:cross_shot}
In addition to the primordial and gravitation contributions, the galaxy bispectrum receives a stochastic contribution due to the discreteness of galaxies. The Poisson prediction for the shot noise contribution is \cite{Jeong:2010,Chan:2016ehg}
\be\label{eq:bis_shot}
B_{\rm g}^{\rm shot}(k_1,k_2,k_3) =   \frac{1}{\bar n_g}\Big(P_g(k_1) + 2 \ {\rm perms}\Big) + \frac{1}{\bar n_g^2}\, ,
\ee
where $\bar n_g$ is the mean number density of the tracer and $P_g(k) = b_1^2 P_0(k)$ denotes the tree-level galaxy power spectrum. The shot noise contribution to the galaxy bispectrum  contributes to the galaxy skew-spectrum as 
\be
P^{\rm shot}_{\tilde D[\delta_g], \delta_g}(k) = \left[ \frac{1}{\bar n_g^2} + \frac{P_g(k)}{\bar n_g} \right] J_{\tilde D}(k) + \frac{2}{\bar n_g}{\tilde J}_{\tilde D}(k)\, , \\
\ee
and
\be
J_{\tilde D}(k) \equiv   \int^\prime_\q \  \tilde D(\q,\k-\q) \, , \quad {\tilde J}_{\tilde D}(k) \equiv   \int^\prime_\q   \tilde D(\q,\k-\q) P_g(q), , 
\ee
where, again, the tilde denotes the product of the ratio of the power spectra and the quadratic kernels, and $\tilde D$ refers to either the gravitational $\tilde D^{\rm grav}$ or primordial $\tilde D^{\rm PNG}$ kernels.

The shot noise contribution for the skew-spectrum should be corrected for exclusion and clustering effects, similarly to corrections to the power spectrum shot noise~\cite{CasasMiranda:2001ym, Baldauf:2013hka}. Following Ref.~\cite{Schmittfull:2014tca}, we model these corrections, for both the gravitational and primordial contributions, with two constant parameters, $\alpha_1$ and   $\alpha_2$, such that 
\be
P^{\rm shot}_{\tilde D[\delta_g], \delta_g }(k) =  \left[\frac{1}{\bar n_g^2}(1+\alpha_2) + \frac{1}{\bar n_g}( 1+\alpha_1)P_g(k) \right] J_{\tilde D}(k) + \frac{2}{\bar n_g}(1+\alpha_1) {\tilde J}_{\tilde D}(k)\, .
\ee
Accounting for $\alpha_2$ as a free parameter in addition to $\alpha_1$ does not significantly improve the fit to $N$-body simulations for Gaussian initial conditions \cite{Schmittfull:2014tca}. In our forecast, we therefore follow their prescription and set  $\alpha_2 = \alpha_1$, and vary $\alpha_1$ as a free parameter.\footnote{Note that $\alpha_1$ in our notation is related to the parameterization of Ref.~\cite{Schmittfull:2014tca} as  $\alpha_1= -A_{\rm shot}$.}

We note that we do not consider a new skew-spectrum for the shot noise contribution. Since we constrain the correction to the Poisson shot noise, it may be a better approach to construct an independent skew-spectrum for this contribution. 

\subsection{Covariance}\label{sec:covs}
In our Fisher analysis, we use the skew-spectra arising from both gravitational evolution and primordial non-Gaussianity to obtain constraints on bias parameters and $\fnl$ of a given shape. We will therefore need the covariance of the gravitational and primordial skew-spectra, as well as their cross-correlations.

At leading-order in perturbation theory,  the covariance of each of the skew-spectra estimators corresponding to primordial and gravitational bispectra is given by \cite{Schmittfull:2014tca}
\begin{align}\label{eq:cov_long}
{\rm Cov} \left[ \hat P_{\tilde D \left[\delta_g\right],\delta_g}(k),\hat P_{\tilde E \left[\delta_g\right],\delta_g}(k')\right] & = \frac{2 (2\pi)^3}{V_i V_s(k)} \tilde P_g(k) I_{\tilde D \tilde E}(k) \delta_{kk'}^{(K)}   + \frac{4}{V_i V_s(k) V_s(k')} \nonumber \\
& \hspace{-2in} \times \int_k \int_{k'}  \d^3 k_1 \d^3k_2 \ \tilde D(\k_1,\k_2-\k_1) \tilde E(\k_2,\k_1-\k_2) \tilde P_g(k_1) \tilde P_g(k_2) \tilde P_g(|\k_1-\k_2|)\, ,
\end{align}
where $\tilde P_g$ denotes the total galaxy power spectrum including the shot noise
\be\label{eq:pg_tilde}
\tilde P_g(k)  = P_g(k) + \frac{1}{\bar n_g}\, ,
\ee
and we defined
\be
I_{\tilde D \tilde E}(k) = \int^\prime_\q \tilde D(\q,\k-\q)\tilde E(\q,\k-\q) \tilde P_g(q) \tilde P_g(|\k-\q|)\, .
\ee
The kernels $\tilde D$ and $\tilde E$ can be either the primordial or gravitational ones given by Eqs.~\eqref{eq:NL_PNG} and \eqref{eq:NL_G}. 
The expression for the off-diagonal contribution to the covariance can be further simplified assuming that the integrand in the second line of Eq.~\eqref{eq:cov_long} is constant over the $k$-bin to obtain\footnote{The second contribution in Eq.~\eqref{eq:full_cov} contributes to both the diagonal and off-diagonal elements of the covariance, and was not evaluated in Ref.~\cite{Schmittfull:2014tca}.} 
\begin{eBox}
\vskip 3pt
\begin{equation}\label{eq:full_cov}
{\rm Cov}_{DE}(k,k') = \frac{2 (2\pi)^3}{V_i V_s(k)} \tilde P_g(k) I_{DE}(k) \delta_{kk'}^{(K)}  + \frac{4 }{V_i }\tilde P_g(k) \tilde P_g(k')  J_{\tilde D \tilde E}(k,k')\, , 
\end{equation}
\vskip 3pt
\end{eBox}
where we defined
\be
J_{\tilde D \tilde E}(k,k') =  \frac{1}{2} \int_{-1}^1 \d\mu \, \tilde D(\k,\k'-\k) \tilde E(\k',\k-\k') \tilde P_g(|\k-\k'|)\, , 
\ee
with $\mu \equiv \hat \k\cdot\hat \k'$. 
Note that both terms in Eq.~\eqref{eq:full_cov} are inversely proportional to the volume of the survey, but their dependence on the volume $V_s$ of $k$-bins is different: while averaging over wider $k$-bin will decrease the contribution of the first term to the covariance, the contribution of the second term can only be reduced by increasing the volume of the survey. These scalings with the $V_i$ and $V_s$ are similar to that of the two (Gaussian vs.~non-Gaussian) contributions to the power spectrum covariance~\cite{Scoccimarro:1999kp}.

\begin{figure}[t]
\centering
\includegraphics[width = 0.85\textwidth]{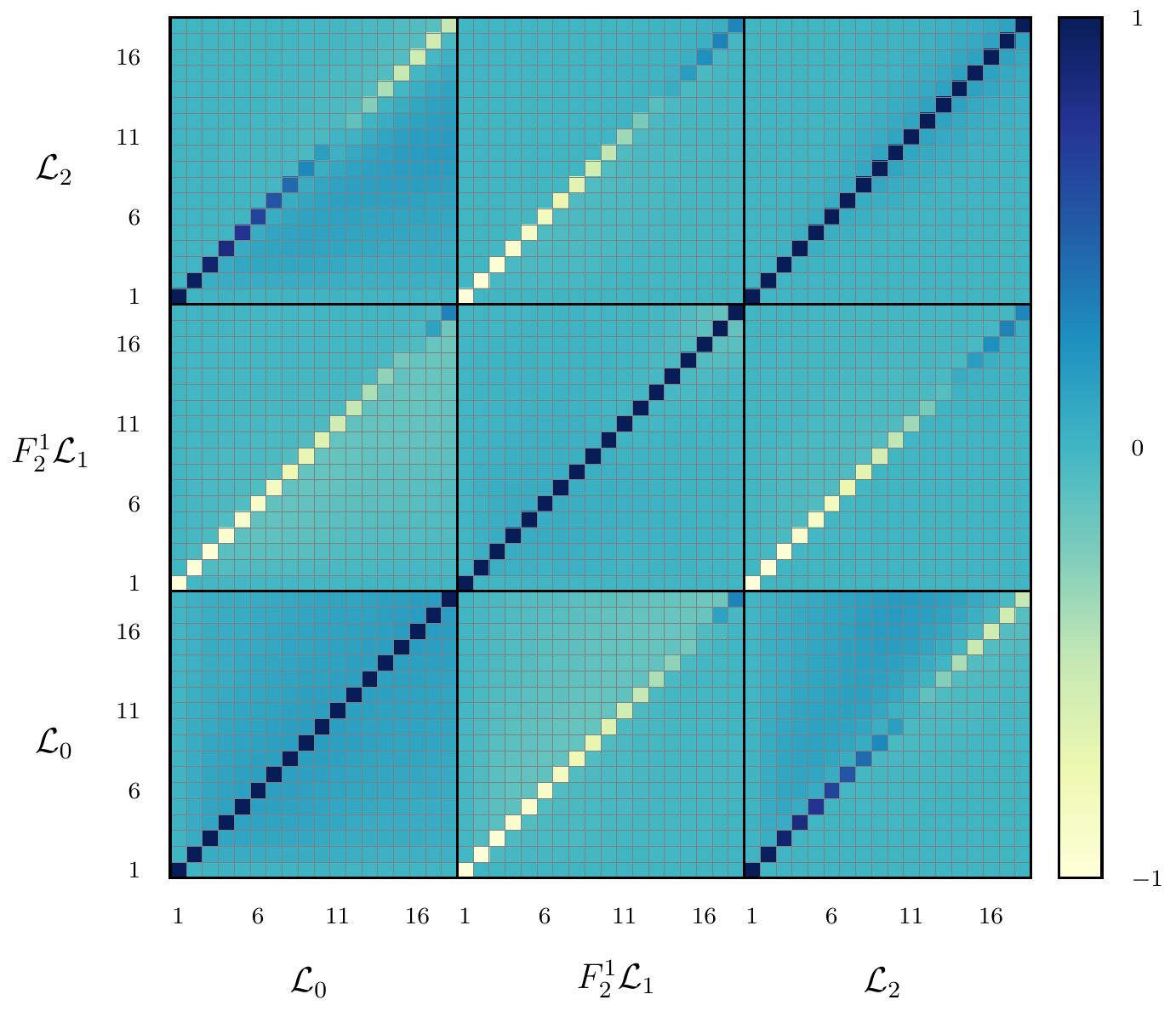}
\caption{The correlation matrix for three skew-spectra corresponding to the nonlinear kernels  $D \in { \{\L_0, F_2^l \L_1, \L_2}\}$. The correlations are calculated at the fifth redshift bin of DESI (LRG+ELG) which centers at $z=1.05$. The $x$ and $y$ axis show the indices of the $k$-bins. For this redshift bin, we have 18 $k$-bins with $\Delta k = 2 k_{\rm min}$. We have set $k_{\rm min} = 0.004 \ {\rm Mpc}^{-1}h$ to be the largest scale corresponding to the volume of the survey in this $z$-bin, and $k_{\rm max} = 0.15 \ {\rm Mpc}^{-1}h$. }
\label{fig:corr_nf1}
\end{figure}

To illustrate the structure of the covariance matrix and the correlations between different $k$-modes for each skew-spectra, in Fig.~\ref{fig:corr_nf1} we show the full correlation matrix for the three skew-spectra corresponding to gravitational evolution. The correlation coefficient between $i$-th and $j$-th $k$-bins is defined as 
\be
r_{ij} = \frac{{\rm Cov}_{ij}}{\sqrt{{\rm Cov}_{ii}{\rm Cov}_{jj}}}\, .
\ee
The correlation matrix has a block structure, with each block being the auto- or cross-correlation of the three skew-spectra. As expected, the off-diagonal elements due to coupling of different $k$-bins in each block are smaller than the diagonals in $k$, and get smaller going away from the diagonal. The diagonal elements corresponding to the correlations of $F_2^1 \mathcal L_1$, with both $\mathcal L_0$ and $\mathcal L_2$ skew-spectra, change sign when going from low to high $k$. For the correlation between the $\mathcal L_0$ and $\mathcal L_2$ spectra, the diagonal has opposite sign, it is positive at lower $k$ and becomes negative at higher $k$.\footnote{The diagonal elements of the correlation matrix agree qualitatively with the results of Ref.~\cite{Schmittfull:2014tca} shown in their Fig.~7. Note that we have opposite signs for the second nonlinear kernel $D = F_2^1 \mathcal L_1$.}

\section{Fisher Forecast} \label{sec:fisher}

In this section, we present Fisher forecasts on the bias parameters, the shot noise correction, and the primordial bispectrum amplitude, from both Gaussian and non-Gaussian initial conditions. In particular, we present the constraints on the parameters due to both the weighted skew-spectra and the full bispectrum, and compare the two results. Our method is described in \S\ref{subsec:method}, and the results are presented in \S\ref{subsec:results}.

\subsection{Methodology}\label{subsec:method}
We perform a Fisher analysis to compare the constraints on the local, equilateral, and spin-exchange primordial bispectra, using the combined skew-spectra and the full bispectrum. In our forecast, we consider the data vector consisting of the three crosses due to gravitational evolution, and the skew-spectrum due to primordial non-Gaussianity in the selected $k$-bins. The total data vector is thus given by 
\be
{\bf d} \equiv \left\{ {\bf P}_{\mathcal L_0 [\delta_g], \delta_g }, \, {\bf P}_{F_2^1 \mathcal L_1 [\delta_g], \delta_g}, \, {\bf P}_{\mathcal L_2[\delta_g], \delta_g }, \, {\bf P}_{ \tilde D^{\rm PNG}[\delta_g],\delta_g} \right\},
\ee
where the first three skew-spectra are due to gravitational evolution, while the last one is due to primordial non-Gaussianity. Each $\bf P$ is the data vector containing the skew-spectra of different $k$-bins. For a given array of parameters  {\boldmath $\lambda$}, the total Fisher matrix of the gravitational and primordial skew-spectra at a given redshift bin with mean $z_i$ is given by 
\be\label{eq:fisher}
F_{\alpha\beta}(z_i) = \sum_{m,n = 1}^{4 N_b} \frac{\partial d_m}{\partial \lambda_\alpha} ({\rm Cov}^{-1})_{mn}  \frac{\partial d_n}{\partial\lambda_\beta}\, ,
\ee
where $N_b$ is the number of $k$-bins with width $\Delta k$, which varies for different redshift bins. Note that the covariance matrix in Eq.~\eqref{eq:fisher} is the full $4N_b\times 4N_b$ covariance matrix for the data vector $\bf d$ defined by
\be
{\rm \bf Cov} \equiv \langle {\bf d} {\bf d} ^t \rangle - \langle {\bf d} \rangle  \langle {\bf d}^t \rangle\, .
\ee
When using the galaxy bispectrum instead of the skew-spectra as the observable, accounting for only the Gaussian contribution to the covariance, the Fisher matrix in each redshift bin is given by 
\be\label{eq:fisher_mbis}
F_{\alpha \beta}(z_i) =  \sum_{k_1\leq k_2\leq k_3= k_{\rm min}}^{k_{\rm max}}  \frac{\partial B_g(k_1,k_2,k_3)}{\partial \lambda_\alpha}  \frac{\partial B_g( k_1,k_2,k_3)}{\partial \lambda_\beta} \, {\rm Var}^{-1} \left[B_g(k_1,k_2,k_3) \right],
\ee
where
\be
{\rm Var}\left[B_g(k_1,k_2,k_3)\right] = \frac{(2\pi)^6 s_{123}}{V_i V_{123}} \   \tilde P_g(k_1)\tilde P_g(k_2)\tilde P_g(k_3)\, ,
\ee
with $s_{123}=6, 2, 1$ for equilateral, isosceles, and scalene triangles, respectively \cite{Scoccimarro:2000sn},  $\bar n_i$ is the mean galaxy number density in the $i$-th redshift bin, $V_{123} = 8 \pi^2 k_1k_2k_3 \Delta k^3$, and $\tilde P_g(k)$ is the galaxy power spectrum including the shot-noise given in Eq.~\eqref{eq:pg_tilde}.

Assuming independent redshift bins, the total Fisher matrix is then the sum of the Fisher matrices over all the redshift bins, 
\be
F_{\alpha \beta} = \sum_i F_{\alpha \beta}(z_i)\, .
\ee
In our main forecast for skew-spectra and bispectrum, we set $k_{\rm min}$ to be the fundamental (largest) mode in the volume of the $i$-th redshift bin, with $\Delta k = k_{\rm min}$ and $k_{\rm max}=0.15 \  h{\rm Mpc}^{-1}$. 

Since the goal of our paper is to show the agreement between constraints derived from the skew-spectra and the bispectrum, rather than providing forecasts from different surveys, we restrict ourselves to  DESI~\cite{Font-Ribera:2013rwa, Aghamousa:2016zmz} as a representative of upcoming spectroscopic surveys. The model of the bispectrum we use does not account for several real-world complications, in particular the redshift-space distortions, the Alock-Paczynski effect, as well as the scale-dependent correction to galaxy bias due to local-shape PNG that is summarized in Appendix \ref{app:full_bis_loc}. In order to make meaningful forecasts for specific surveys, one needs to account for all of these. We leave more complete forecasts for future work. 

We use the same survey designs as in our previous work~\cite{MoradinezhadDizgah:2018ssw}. Here we only briefly outline our choices and refer the reader to more details to that reference. We assume uncorrelated top-hat redshift bins, such that the shot-noise in the $i$-th redshift bin is given by 
\be
\bar n_i = \frac{4\pi f_{\rm sky}}{V_i} \int_{z_{\rm min}}^{z_{\rm max}} \d z\,  \frac{\d N}{\d z}(z)\,,
\ee
where $f_{\rm sky}$ is the fraction of the sky covered, and $\d N/\d z$ is the surface number density of galaxies in each survey (see Fig.~5 in Ref.~\cite{MoradinezhadDizgah:2018ssw}). We take $f_{\rm sky} = 0.34$ corresponding to a coverage of  $14000 \ {\rm deg}^2$, and consider the 12 redshift bins in the range of $0.65\leq z \leq 1.65$.

We fix the cosmological parameters and only vary the bias parameters, as well as the correction to the shot noise, in addition to the amplitudes of the primordial bispectra when considering non-Gaussian initial conditions. 
We set the values of the cosmological parameters to the best-fit values of Planck 2018~\cite{Aghanim:2018eyx} (TT,EE,TE+lowE+lensing+BAO): $\ln(10^{10}A_s) = 3.047, n_s =  0.9665, h = 0.6766, \Omega_c h^2= 0.11933, \Omega_b h^2=   0.02242, \tau =0.0561$, with pivot scale $k_p= 0.05\ {\rm Mpc}^{-1}$. 
We use the public CLASS code \cite{Lesgourgues:2011re,Blas:2011rf} to compute the linear matter power spectrum. 
For non-Gaussian initial conditions, we set the fiducial value of the amplitude to $f_{\rm NL}= 0$. 

We model the redshift evolution of the linear bias as $b_1(z) = \bar b_1 p(z)$, where $\bar b_1$ is a free amplitude that we vary and $p(z)$ captures the redshift dependence. We set the fiducial value of $\bar b_1 = 1.46$ such that at $z=0$ the value of the linear bias is consistent with the results in Ref.~\cite{Lazeyras:2015lgp} for halos of mass $M = 3\times 10^{13} h^{-1} M_\odot$. 
We take $p(z) = 0.84/D(z)$ where $D(z)$ is the growth factor normalized to unity at $z=0$~\cite{Font-Ribera:2013rwa}. For the fiducial values of quadratic biases we assume the scaling relations of $b_2 = \bar b_2(0.412 -  2.143 b_1 + 0.929 b_1^2 + 0.008 b_1^3)$ and $b_{K^2} = \bar b_{K^2}(0.64 - 0.3 b_1 + 0.05 b_1^2 -0.06 b_1^3)$, which are fits to $N$-body simulations~\cite{Lazeyras:2015lgp,Modi:2016dah}. Based on these results, we assume that the above relation between $b_2$ and $b_{K^2}$ with $b_1$, is preserved in the redshift range we consider, and use it to set the fiducial values of the the biases in each redshift bin.
In the Fisher forecast, we then vary two parameters for the overall amplitudes $\bar b_2$ and $\bar b_{K^2}$. The fiducial value of the correction to the shot-noise is set to  $\alpha = 0$.

The parameter array in our forecast for Gaussian initial conditions is then {\boldmath $\lambda$} $= \{\bar b_1,\bar b_2,\bar b_{K^2},\alpha\}$, while when accounting for primordial non-Gaussianity we also vary $f_{\rm NL}$. We note that since the skew-spectra derived here are only optimal estimators for parameters appearing as amplitudes, in the case of the primordial bispectrum due to massive particles with spin, we fix the mass parameter such that $\nu=3$. In our previous studies on the observablity of this shape of non-Gaussianity using the galaxy bispectrum \cite{MoradinezhadDizgah:2018ssw}, we also varied the mass parameter $\nu$.  

\subsection{Results}\label{subsec:results}
In this section, we first present  forecasted constraints on the halo bias parameters $b_1,b_2,b_{K^2}$, as well as the shot noise correction $\alpha$ for Gaussian initial conditions. When considering non-Gaussian initial conditions, in addition to the above three parameters, we also vary the $\fnl$ of the local, equilateral, and spin-2 exchange bispectra. 

\subsubsection{Gaussian Initial Conditions}\label{sec:G_ICs}
\begin{figure}[t]
\begin{center}
\begin{tabular}{| p{40mm} || >{\centering\arraybackslash}  m{22mm}  | >{\centering\arraybackslash} m{22mm} | >{\centering\arraybackslash}  m{22mm}  | >{\centering\arraybackslash} m{22mm} |}
 \hline
\hspace{.4in}  DESI & $\sigma(\bar b_1)$ & $\sigma(\bar b_2)$  & $\sigma(\bar b_{K^2})$ &   $\sigma(\alpha)$  \\
 \hline
 \hline
Bispectrum  &  $  4.37 \times 10^{-3}$ &  $  7.29 \times 10^{-3}$ & $ 1.91 \times 10^{-2}$   &  $  1.72 \times 10^{-3}$ \\
Skew-spectra, full cov    & $  4.13 \times 10^{-3}$ &  $  7.24 \times 10^{-3}$ & $ 1.71 \times 10^{-2}$   &  $  1.94 \times 10^{-3}$ \\
Skew-spectra, diag cov  & $  2.60 \times 10^{-3}$ &  $  4.44 \times 10^{-3}$ & $ 1.51 \times 10^{-2}$   &  $  1.25 \times 10^{-3}$ \\
 \hline
\end{tabular}
\captionof{table}{1-$\sigma$ marginalized constraints on the bias parameters, the shot-noise correction and amplitude of primordial bispectrum, obtained from the galaxy bispectrum and skew-spectra with and without off-diagonal elements for the DESI survey.}\label{tab:G_1sig}
\vspace{.3in}
\includegraphics[width= 0.496\textwidth,height= 0.36\textheight]{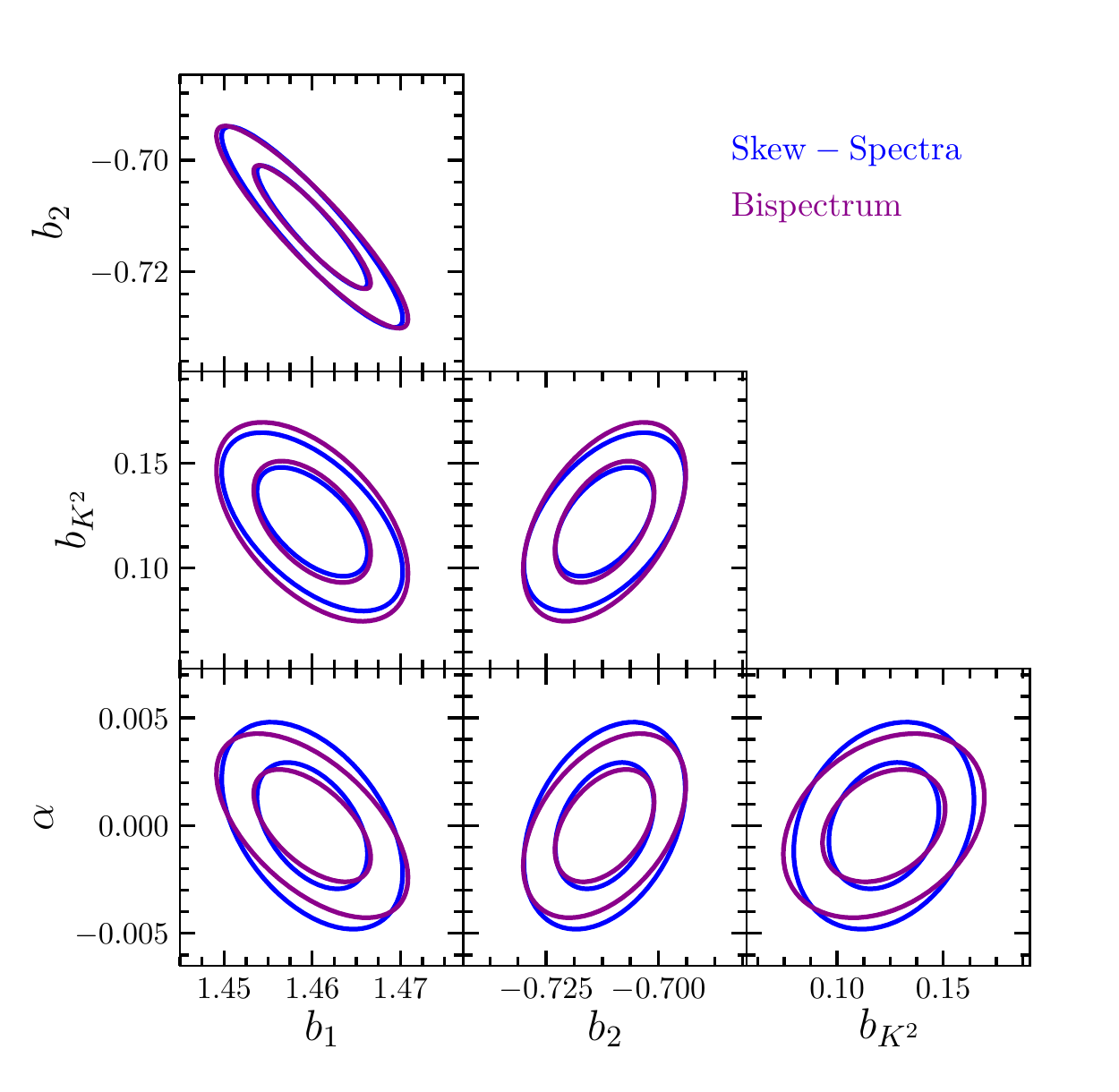}
\includegraphics[width= 0.496\textwidth,height= 0.36\textheight]{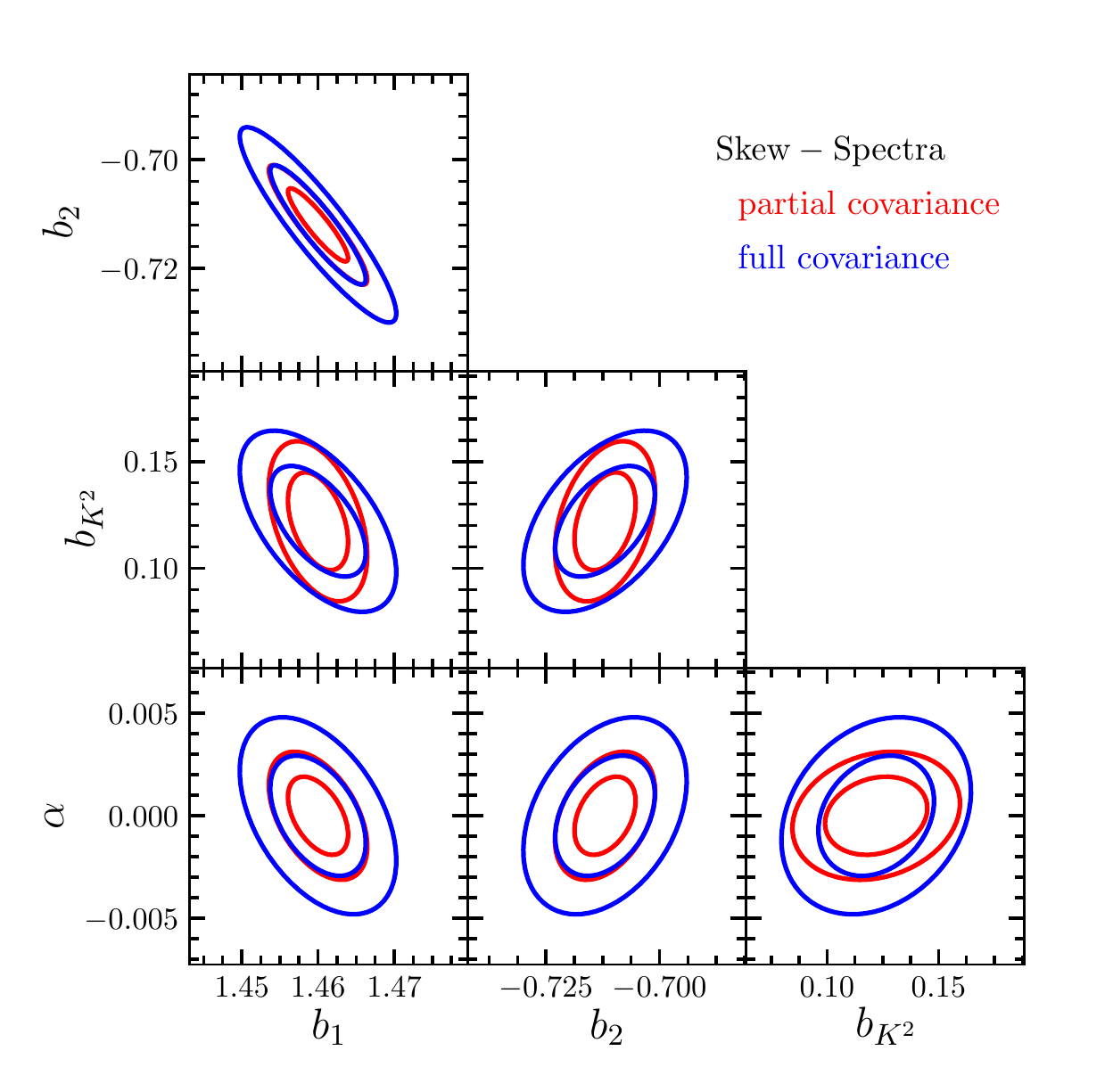}
\captionof{figure}{Forecast for two-dimensional marginalized constraints on bias parameters and the shot-noise correction.  The left panel shows the constraints from the skew-spectra using their full covariance (blue), and for the  bispectrum (purple). The right panel shows that constraints for the skew-spectra would be overly optimistic when assuming a diagonal covariance (red) instead of the full covariance (blue).
The fiducial values of the parameters are given in the text, and the forecast is for the DESI survey.}
\label{fig:G_cross}
\end{center}\vspace{-.2in}
\end{figure}

In Table~\ref{tab:G_1sig}, we show the forecasted one-dimensional 1-$\sigma$ marginalized constraint on each parameter, while in Fig.~\ref{fig:G_cross} we show the forecasted two-dimensional marginalized constraints on each pair of parameters. We compare the constraints obtained from the bispectrum, the skew spectra using the full covariance, and the skew spectra using only the first term in Eq.~\eqref{eq:full_cov} for the covariance. The first two rows in Table \ref{tab:G_1sig} and the left panel of Fig.~\ref{fig:G_cross} show that the constraints from the skew-spectra and the bispectrum are comparable within a few percent if the full covariance of the skew-spectra is taken into account. This shows numerically that the skew-spectra capture the full information of the bispectrum on amplitude-like parameters, i.e.~biases and the shot-noise correction. The bottom line of Table~\ref{tab:G_1sig} and the right panel of Fig.~\ref{fig:G_cross}  show  that errors on the parameters are severely underestimated and the degeneracy direction of some parameters is tilted compared to using the full covariance or the bispectrum. This highlights the fact that it is  crucial to include the full covariance of the skew-spectra to obtain correct constraints. \\

\subsubsection{Non-Gaussian Initial Conditions}\label{sec:NG_ICs}
In Table~\ref{tab:PNG} we show the forecasted 1-$\sigma$ constraints on the three primordial bispectrum shapes that we consider. The constraints are overall in good agreement with each other, with a mild discrepancy of a few percent. As in the case of Gaussian initial conditions, this difference is partially driven by the fact that we do not introduce a new skew-spectrum corresponding to the shot-noise contribution to the bispectrum.  This results in weaker constraints on $\alpha$ from the skew-spectrum compared to that from the bispectrum. An additional source of error is the numerical precision in calculating the skew-spectra and their covariances.

\begin{figure}
\begin{center}
\hspace{-.14in}\includegraphics[width=0.49\textwidth,height= 0.35\textheight]{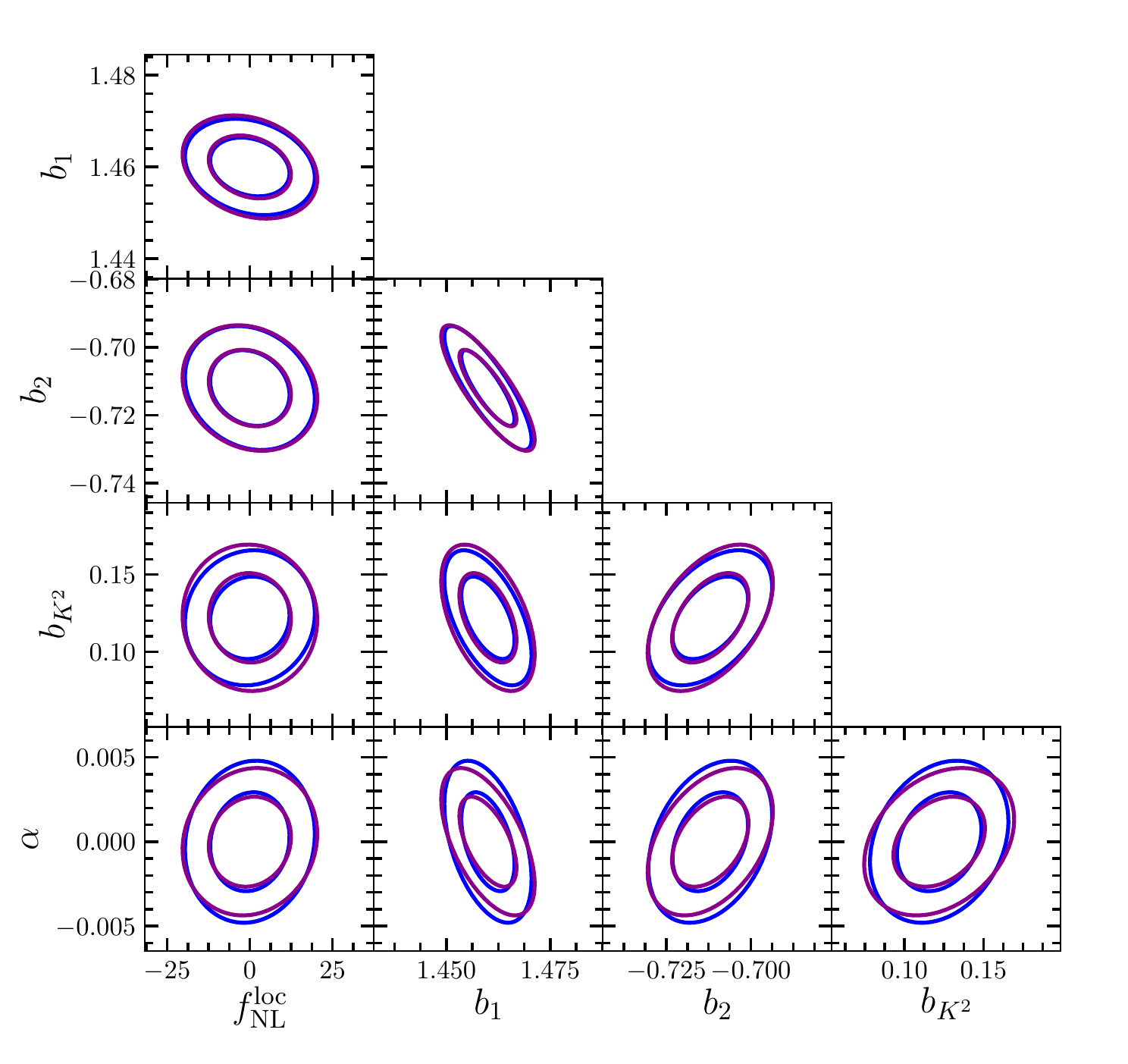}
\hspace{.1in}\includegraphics[width= 0.49\textwidth,height= 0.35\textheight]{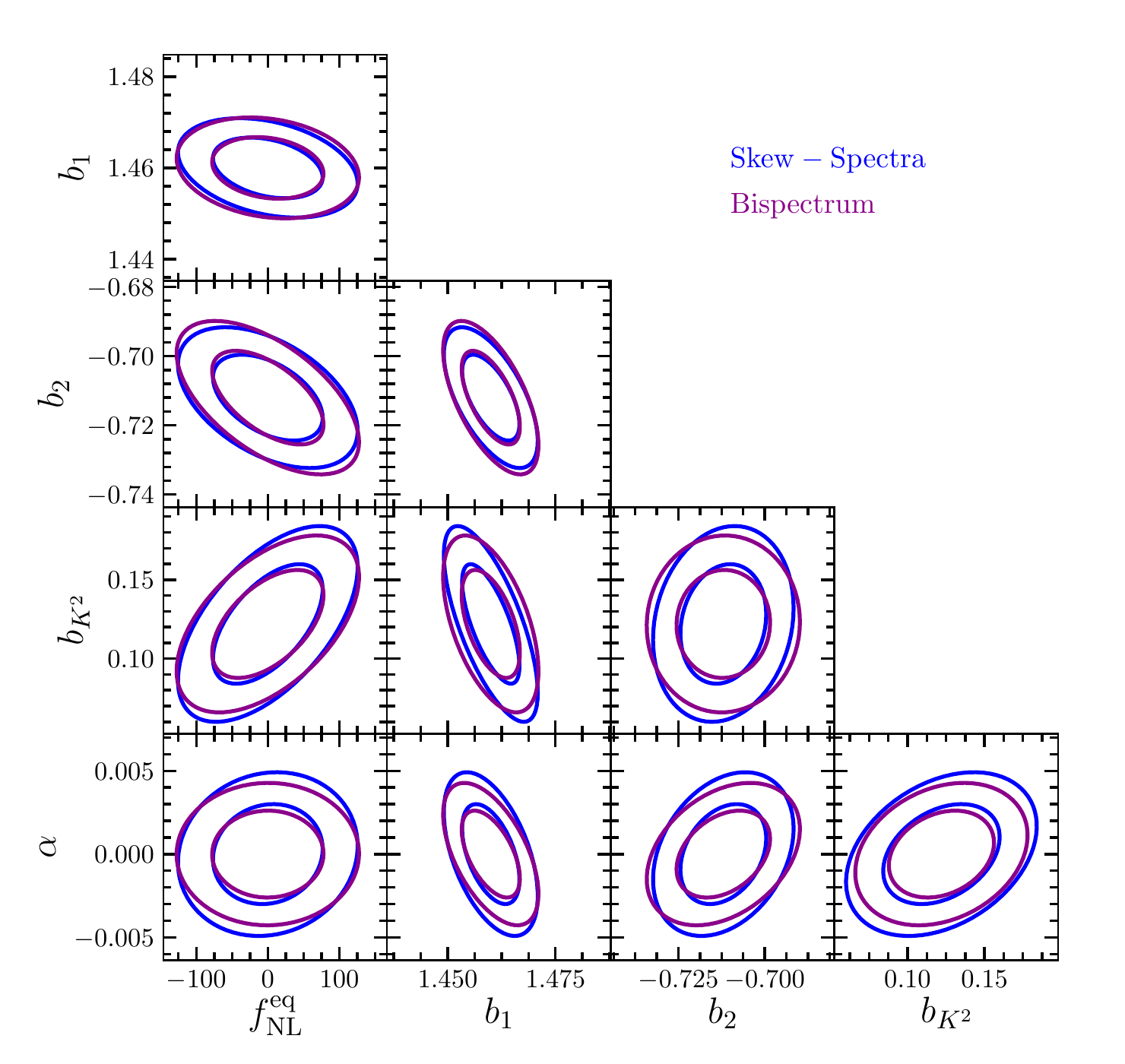}
\end{center}
\vspace{-.1in}
\hspace{-.14in}\includegraphics[width= 0.49\textwidth,height= 0.35\textheight]{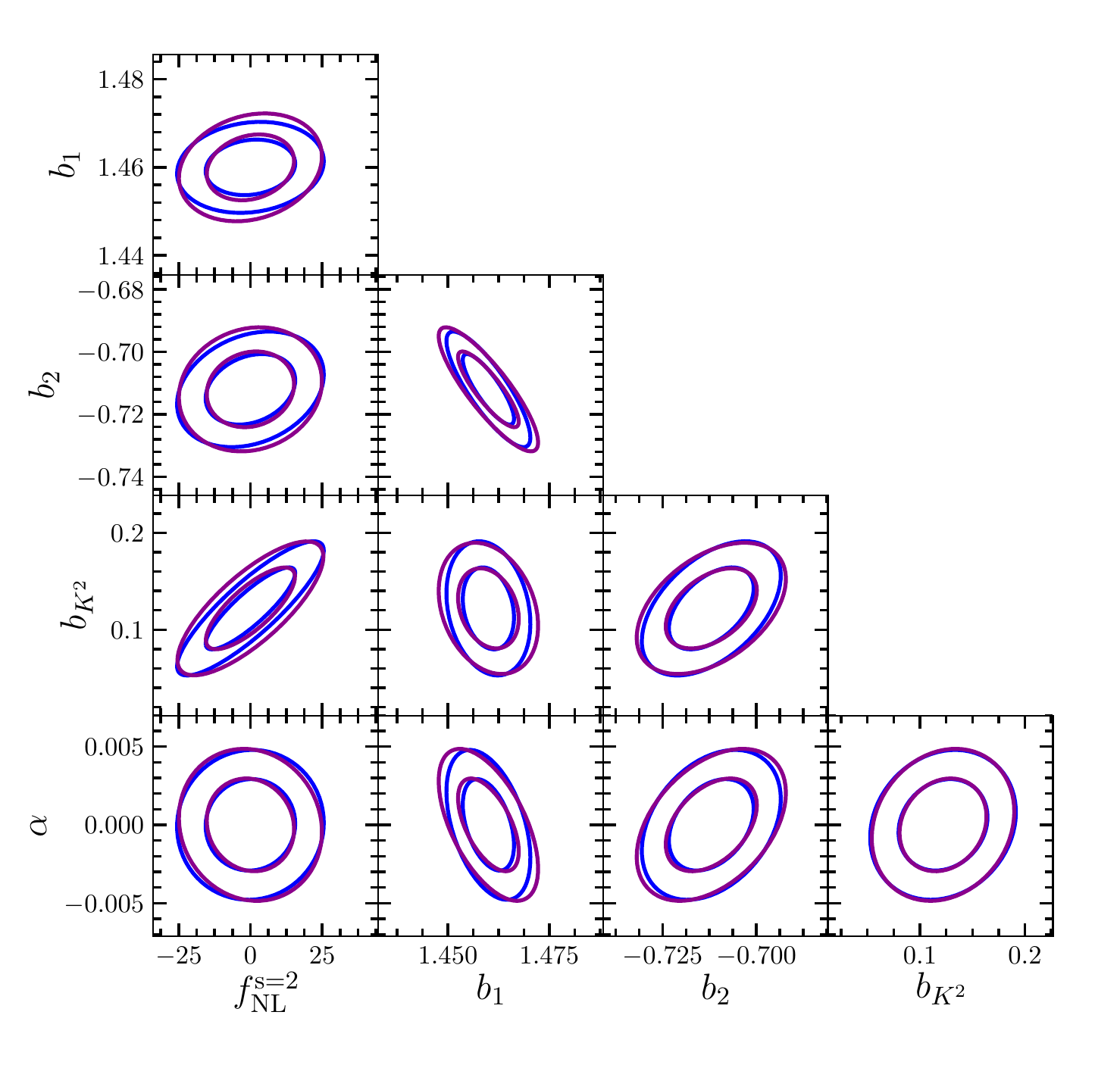}
\hfill
\begin{minipage}{.45\textwidth}
\vspace{-2.4in}
\caption{Forecasted two-dimensional marginalized constraints on the bias parameters, the shot-noise correction and amplitude of primordial bispectrum of the local  (top left panel) and equilateral (top right panel) shapes, as well as that due to massive particles with spin $s=2$ (bottom panel). The blue (purple) lines correspond to skew-spectra (bispectrum). The fiducial values of the parameters are given in the text, and the constraints are obtained for the DESI survey.}
\label{fig:PNG_loc_eq_s2}
\end{minipage}\vspace{-.8in}
\end{figure}

In Fig.~\ref{fig:PNG_loc_eq_s2}, we show the two-dimentional constraints on each pair of the parameters, for the local shape on the left and equilateral shape on the right. The blue contours are constrains from the skew-spectra while the purple contours are obtained from the bispectrum. As for the Gaussian initial conditions, the constraints agree with each other to a good approximation. 

\begin{table*}
\begin{tabular}{| p{25mm} || >{\centering\arraybackslash}  m{22mm}  | >{\centering\arraybackslash}  m{22mm}  |>{\centering\arraybackslash} m{22mm} | >{\centering\arraybackslash}  m{22mm}  | >{\centering\arraybackslash} m{22mm} |}
 \hline
\hspace{.3in}  DESI & $ \sigma(f_{\rm NL}^{\rm loc})$ & $\sigma(\bar b_1)$ & $\sigma(\bar b_2)$  & $\sigma(\bar b_{K^2})$ &   $\sigma(\alpha)$  \\
 \hline
 \hline
Bispectrum  & 8.15  &  $  4.50 \times 10^{-3}$ &  $  7.29\times 10^{-3}$ & $ 1.91 \times 10^{-2} $ & $ 1.73 \times 10^{-3}$ \\
Skew-spectra    & 7.91  & $4.23 \times 10^{-3}$ &   $7.37 \times 10^{-3}$ &  $1.76  \times 10^{-2} $& $1. 93  \times 10^{-3}$ \\ 
 \hline 
\end{tabular}\vspace{.07in}  
\begin{tabular}{| p{25mm} || >{\centering\arraybackslash}  m{22mm}  | >{\centering\arraybackslash}  m{22mm}  |>{\centering\arraybackslash} m{22mm} | >{\centering\arraybackslash}  m{22mm}  | >{\centering\arraybackslash} m{22mm} |}
 \hline
\hspace{.3in}  DESI & $ \sigma(f_{\rm NL}^{\rm eq})$ & $\sigma(\bar b_1)$ & $\sigma(\bar b_2)$  & $\sigma(\bar b_{K^2})$ &   $\sigma(\alpha)$  \\
 \hline
 \hline
Bispectrum  & 51.3&  $4.45 \times 10^{-3}$ &  $8.93 \times 10^{-3}$&  $ 2.26 \times 10^{-2}$&  $1.72 \times 10^{-3}$\\
Skew-spectra   & 50.6 & $  4.39 \times 10^{-3}$ &  $ 8.19 \times 10^{-3}$&  $ 2.50 \times 10^{-2}$&  $1.98  \times 10^{-3}$ \\
 \hline
\end{tabular}\vspace{0.07in}
\begin{tabular}{| p{25mm} || >{\centering\arraybackslash}  m{22mm}  |  >{\centering\arraybackslash}  m{22mm}  |>{\centering\arraybackslash} m{22mm} | >{\centering\arraybackslash}  m{22mm}  | >{\centering\arraybackslash} m{22mm} |}
 \hline 
\hspace{.3in}  DESI & $ \sigma(f_{\rm NL}^{s=2})$ & $\sigma(\bar b_1)$ & $\sigma(\bar b_2)$  & $\sigma(\bar b_{K^2})$ &   $\sigma(\alpha)$  \\
 \hline
 \hline
Bispectrum  & 10.0 &  $4.93 \times 10^{-3}$  &  $8.01 \times 10^{-3}$&  $2.73 \times 10^{-2}$&  $1.95 \times 10^{-3}$\\
Skew-spectra   & 10.0 &  $4.17 \times 10^{-3}$& $ 7.46 \times 10^{-3}$ &  $2.79 \times 10^{-2}$ &  $1.93 \times 10^{-3}$ \\
 \hline
\end{tabular}
\caption{Forecasted 1-$\sigma$ marginalized constraints on the bias parameters, the shot-noise correction and the amplitude of primordial bispectra of the local, equilateral shapes, as well as the bispectrum due to the presence of massive particles with spin $s=2$.}
\label{tab:PNG}
\end{table*}

\newpage
\section{Conclusions}\label{sec:con}
Higher-order clustering statistics of galaxies contain a wealth of information on cosmology. On the one hand, they can be used to test the modeling of  nonlinearities in the power spectrum and to improve  constraints by breaking parameter degeneracies. On the other hand, they provide information not captured by the power spectrum, such as the imprints of primordial non-Gaussianity beyond the local shape. Up to now, most of the cosmological information from LSS surveys comes from measurements of the power spectrum. However, the next generation of galaxy surveys will provide high precision measurements of the bispectrum, thanks to the large cosmological volume of the surveys and the high number density of the observed galaxies. 

In this paper, we investigated the information content of weighted skew-spectra \cite{Schmittfull:2014tca} as proxies for the galaxy bispectrum. 
Using a Fisher forecast, we compared the expected constraints on galaxy bias parameters, corrections to Poisson shot noise and the amplitude of primordial non-Gaussianity from the full galaxy bispectrum against skew-spectra. 
We find that the skew-spectra have the same Fisher information for these amplitude-like parameters, i.e.~they fully capture the information content of the bispectrum on these parameters. To obtain this agreement, it is crucial to account for the full covariance matrix of the skew-spectra, including mode-coupling contributions that were previously neglected.

The skew-spectra are constructed to be cross-spectra between the observed density field and a configuration-space product of two appropriately filtered density fields. The filters derive from the shape of the bispectrum contribution under consideration, and the form of the estimator derives from the maximum likelihood estimator for the bispectrum amplitude in the limit of weak non-Gaussianity. 
In Fourier space, the product corresponds to a convolution of the two filtered fields, and integrating the cross-spectrum over all wavenumbers  corresponds to summing over all bispectrum triangle configurations. For Gaussian initial conditions, three skew-spectra constructed from the matter density, the displacement and the tidal tensor, $\delta_m^2, \psi^i\partial_i\delta_m, [K_{ij}]^2$, are needed, while in the presence of primordial non-Gaussianity, one additional quadratic field constructed from the gravitational potential at the initial conditions is needed. 

In order to apply the skew-spectra to actual observational data as a proxy for the bispectrum, one needs to account for several additional ingredients. First, to compute the theoretical skew-spectra, we performed all integrals numerically. In order to embed the skew-spectra in Monte Carlo Markov Chain parameter estimation, the computation of these integrals should be accelerated. Given the similarity of these integrals with mode-coupling integrals in the loop computation of the power spectrum and bispectrum of the LSS, the results of recent studies~\cite{McEwen:2016fjn,Schmittfull:2016jsw,Simonovic:2017mhp} should be applicable to achieve the required speed-up.  Alternatively, one could build an emulator for the skew-spectra using simulations of different cosmologies, which exist already for building power spectrum emulators. 

Also, the redshift-space distortions (RSD) and the Alcock-Paczynski (AP) effects were not modeled. To account for RSD, several new skew-spectra are needed, while modeling of AP entails a re-scaling of the wavenumbers and angles in the skew-spectra. We do not expect any conceptual difficulty in implementing these effects, although there may be numerical challenges, especially in modeling of the RSD. This is because the data vector enlarges significantly due to the additional skew-spectra. 

We only accounted for the the skew-spectra corresponding to the tree-level bispectrum. In order to use the information on small scales, modeling the bispectrum beyond the tree-level is necessary. Unlike the tree-level bispectrum, the 1-loop bispectrum does not have a separable form. This issue can potentially be overcome by realizing that the shapes of 1-loop contributions are highly correlated with the tree-level shape~\cite{Lazanu:2015rta}, which implies that measuring the skew-spectra that are derived from the tree-level bispectrum will capture most of the information in the full bispectrum. However, in that case, one has to be more careful in interpreting the measured amplitude parameters. Also, we did not fully account for contributions to biases from primordial non-Gaussianity. As we discussed earlier and describe in Appendix~\ref{app:full_bis_loc}, additional skew-spectra should be introduced for local non-Gaussianity to capture the contributions of the (scale-dependent) corrections to galaxy biases. 

Lastly, we restricted the analysis to parameters that rescale the amplitude of bispectrum contributions. It would be interesting to see if optimal skew-spectra can also be constructed for cosmological parameters that modify the shape of the bispectrum. This might be possible by modeling the response of the bispectrum to such parameters with separable templates.

\acknowledgments
We thank Tobias Baldauf, Joyce Byun, Kwan Chuen Chan, Vincent Desjacques, Airam Marcos-Caballero, Lado Samushia and Emiliano Sefusatti for very helpful discussions and insights. We also thank Emiliano Sefusatti for providing feedback on the manuscript. A.M.D.~is supported by the SNSF project, {\it The Non-Gaussian Universe and Cosmological Symmetries}, project number: 200020-178787. HL and CD are supported by Department of Energy (DOE) grant DE-SC0020223. MS gratefully acknowledges support from the Corning Glass Works Fellowship and the National Science Foundation.

\appendix

\section{Galaxy Bispectrum from Local Non-Gaussianity}\label{app:full_bis_loc}
The nonzero primordial bispectrum has an additional contribution to the bispectrum of biased tracers. For instance, local non-Gaussianity leads to an additional effect on the galaxy density field, so that it now depends on the gravitational potential $\phi$ at early times, as well as on the matter density field. This means that in the presence of local non-Gaussianity we need to include additional bias coefficients in the bias expansion. Keeping only terms linear in the gravitational potential $\phi$ and quadratic in the density and tidal shear, and neglecting the stochastic terms as well as the higher-derivative terms of $\delta$ and $\phi$, the bias expansion is given by
\begin{align}\label{eq:bias_png}
\delta_g(\x) &= b_1 \delta_m(\x) + b_2 \delta_m(\x)^2 +  b_{K^2}\left[K_{ij}(\x)\right]^2 \nonumber \\
&+ f_{\rm NL}^{\rm loc} \left[ b_\phi \phi(\q) + b_{\phi \delta}\phi(\q) \delta(\x) + b_{\phi \delta^2}\phi(\q) \delta^2(\x) + b_{\phi K^2}\phi(\q)(K_{ij})^2(\x) \right].
\end{align}
The first $f_{\rm NL}$-dependent term accounts for the fact that primordial non-Gaussianity of local shape changes the amplitude (or variance) of the small-scale fluctuations. Therefore, the value of this bias can be obtained by the response of the mean number density of galaxies to a rescaling of the amplitude. Assuming a universal mass function for halos, this gives us the known result for the scale-dependant bias due to local-shape primordial bispectrum from the peak-background split (PBS) argument:

\be
\Delta b_1(k,z) = f_{\rm NL}^{\rm loc} b_\phi \mathcal M^{-1}(k,z) = \frac{6}{5}f_{\rm NL}^{\rm loc} \delta_c (b_1-1)\mathcal M^{-1}(k,z)\, .
\ee
The PBS can be further applied to obtain the amplitude of the higher-order non-Gaussian biases in Lagrangian space \cite{Giannantonio:2009ak}
\be
b_{\phi\delta}^L = -b_1^L  + \delta_c b_2^L\, ,
\ee
which then can be related to the Eulerian ones
\be
b_{\phi \delta} = b_{\phi \delta}^L + b_\phi\, .
\ee
Here $\delta_{\rm c} =1.686 $ is the critical threshold for spherical collapse of halos. The second-order Lagrangian bias is related to Eulerian biases as $b_2^L = b_2 - 8/21(b_1-1)$. 

Using the bias expansion in Eq.~\eqref{eq:bias_png} and keeping terms linear in $f_{\rm NL}^{\rm loc}$, the galaxy bispectrum in the presence of local bispectrum is then given by~\cite{Desjacques:2016bnm}
\begin{align}\label{eq:loc_bis}
B_g(k_1,k_2,k_3) &= B_{\rm g}^{\rm G}(k_1,k_2,k_3) + B^{\rm PNG}_{\rm g}(k_1,k_2,k_3) + \bigg\{ b_1^2 \bigg[\mu_{12} \left(\frac{k_1}{k_2}\Delta b_1(k_1) + \frac{k_2}{k_1} \Delta b_1(k_2) \right)  \nonumber \\
 & + \frac{(2\delta_c-1)(b_1-1) + \delta_c b_2^L}{2\delta_c(b_1-1)}\left[\Delta b_1(k_1) + \Delta b_1(k_2)\right] \bigg]P_0(k_1)P_0(k_2) \nonumber \\
& + 2 b_1^2 \left[\Delta b_1(k_1) + \Delta b_1(k_2) + \Delta b_1(k_3) \right]  F_2(\k_1,\k_2) P_0(k_1)P_0(k_2)\nonumber \\
&  + 2 b_1 \left[\Delta b_1(k_1) + \Delta b_1(k_2) \right] \left[ b_2+b_{K^2}\left(\mu_{12}^2 -\frac{1}{3}\right)\right] P_0(k_1)P_0(k_2)  + 2 \ {\rm perm}\bigg\},
\end{align}
where $B_g^{\rm PNG}$ is the direct contribution of the primordial bispectrum to the galaxy bispectrum
\be
B^{\rm PNG}_{\rm g}(k_1,k_2,k_3) =  b_1^3 \ {\mathcal M}(k_1){\mathcal M}(k_2) {\mathcal M}(k_3)  B_\zeta(k_1,k_2,k_3)\, .
\ee
Note that, as was shown in Ref.~\cite{MoradinezhadDizgah:2017szk}, for the primordial bispectrum due to the presence of massive particles with spin, $b_\phi$ is suppressed and has a weak to no scale-dependance (depending on the spin of particles). In our forecast for the bispectrum and skew-spectra, we therefore only keep the first two terms in Eq.~\eqref{eq:loc_bis}, and leave the analysis of the full bispectrum for future work.

\section{Separable Templates}\label{sec:sep_temp}
All physical primordial correlators have certain powers of $k_t$ appearing in denominators. One way to make this separable is to use the Schwinger parameterization and then approximate the integral as a finite sum, as described in the main text. However, it is often possible to directly construct a separable template that approximates the original shape to high precision. In this appendix, we describe a procedure for constructing separable templates for the analytic part of the spin-exchange bispectra.

First, in order to see how much the template is correlated with the original shape, it is convenient to introduce the dimensionless shape function
\begin{align}
	S(k_1,k_2,k_3) = \frac{(k_1k_2k_3)^2}{A_s^2}B(k_1,k_2,k_3)\, ,
\end{align}
and use the following inner product between two shape functions~\cite{Babich:2004gb}
\begin{align}
	\langle S_a,S_b\rangle =\int_0^1\d x\int_{1-x}^1 \d y\, S_a(1,x,y)S_b(1,x,y)\, .\label{inner}
\end{align}
The goal is to find a simple separable template that is highly correlated with the original shape.

The analytic part of the bispectrum is generated from the interaction $\dot\pi(\hat\partial_{i_1\cdots i_s}\pi)^2$, where $\hat\partial_{i_1\cdots i_s}$ denotes the traceless part of $\partial_{i_1\cdots i_s}$. Assuming scale invariance, the bispectrum takes the form
\begin{equation}
	B^{\rm A}_s(k_1,k_2,k_3) = \frac{\hskip -0.5pt\L_s(\hat \k_1\cdot\hat \k_2)}{ (k_1k_2)^{3-s}k_3 k_t^{2s+1}}\Big[(2s-1)\big((k_1+k_2)k_t+2sk_1k_2\big)+k_t^2 \Big] + \text{2 perms}\, .\label{Banalytic}
\end{equation}
This is not separable due to the $k_t$ factor appearing in the denominator. 
A manifestly factorizable template should be of the form
\begin{align}
	B^{\rm temp}_s(k_1,k_2,k_3)= \frac{Q_s(k_1,k_2,k_3)}{(k_1k_2k_3)^p}\, ,
\end{align}
where $Q_s(k_1,k_2,k_3)$ is a polynomial that is symmetric in its arguments and $p=2+{\rm deg}[Q_s]$. There are a number of existing methods for building a basis functions for $Q_s$, e.g.~\cite{Fergusson:2009nv, Byun:2013jba, Byun:2015rda}. Instead, let us consider an ansatz given by
\begin{equation}
\hskip -7pt	 B_{n_1n_2n_3}^{p_1p_2p_3}(k_1,k_2,k_3)= \frac{(k_2^{p_1}+k_3^{p_1}-k_1^{p_1})^{n_1}(k_3^{p_2}+k_1^{p_2}-k_2^{p_2})^{n_2}(k_1^{p_3}+k_2^{p_3}-k_3^{p_3})^{n_3}}{(k_1k_2k_3)^{2-\frac{1}{3}(n_1p_1+n_2p_2+n_3p_3)}}+\text{2 perms}\, ,\label{Bansatz}
\end{equation}
with integer $n_i$, $p_i$. 
This ansatz is motivated by the shape dependence of the original bispectrum \eqref{Banalytic}. For instance, the shape involves symmetrizing the Legendre polynomial; for $s=1,2$, this gives
\begin{align}
	\L_1(\hat\k_1\cdot\hat\k_2) + \text{2 perms} 
	\, & \propto\, \frac{(k_2+k_3-k_1)(k_3+k_1-k_2)(k_1+k_2-k_3)}{k_1k_2k_3}+2\, ,\label{P1sym}\\[5pt]
	\L_2(\hat\k_1\cdot\hat\k_2) + \text{2 perms} 
	\, & \propto\, \frac{(k_2^2+k_3^2-k_1^2)(k_3^2+k_1^2-k_2^2)(k_1^2+k_2^2-k_3^2)}{(k_1k_2k_3)^2}\, .\label{P2sym}
\end{align}
The terms in the square brackets in Eq.~\eqref{Banalytic} is monotonic, so most of the shape dependence comes from the Legendre polynomials. This suggests that Eq.~\eqref{Bansatz} forms a natural basis that we can use. 
From this building block, we construct a general separable bispectrum by
\begin{align}
	B^{\rm temp}_s(k_1,k_2,k_3) = \sum_{n_i,p_i}a_{n_1n_2n_3}^{p_1p_2p_3} B_{n_1n_2n_3}^{p_1p_2p_3}(k_1,k_2,k_3)\, ,
\end{align}
with coefficients $a_{n_1n_2n_3}^{p_1p_2p_3}$. 
The function $Q_s$ is constrained by demanding that $B_s^{\rm temp}$ has the correct scaling behavior in the squeezed limit: Eq.~\eqref{Bangle} behaves as $k_1^{-1}$ as $k_1\to 0$. The leading squeezed limit of \eqref{Bansatz} is given by $B_{n_1n_2n_3}^{p_1p_2p_3}(q\to 0,k,k) \sim q^w$, where
\begin{align}
	w = -2-\frac{1}{3}(n_1p_1+n_2p_2+n_3p_3)+\text{min}(n_1p_1+n_2p_2,n_1p_1+n_3p_3,n_2p_2+n_3p_3)\, .
\end{align}
It turns out that the unique combination that lead to $w=-1$ is given by $n_i=p_i=1$; this is nothing other than the usual equilateral template
\begin{align}
	B^{\rm eq} = \frac{(k_1+k_2-k_3)(k_2+k_3-k_1)(k_3+k_1-k_2)}{(k_1k_2k_3)^3}\, . \label{Beq}
\end{align}
This makes sense, since the equilateral template approximates the shape that arises from $\dot\pi(\partial_i\pi)^2$, which is nothing but Eq.~\eqref{Banalytic} with $s=1$. The fact that this is unique implies that any higher-spin template must also contain the equilateral template. 
For general spin, we expect the degree of the polynomial to match Eq.~\eqref{P2sym}; e.g.~we want $n_1p_1+n_2p_2+n_3p_3=6$ for $s=2$. There are four possible choices: $B_{111}^{222}$, $B_{112}^{221}$, $B_{122}^{211}$, $B_{222}^{111}$. 
It turns out to be sufficient to have just one of these, and we find that
\begin{align}
	B_{s=2}^{\rm temp}(k_1,k_2,k_3) = B^{\rm eq}(k_1,k_2,k_3)+\frac{1}{5}B_{112}^{221}(k_1,k_2,k_3)\, ,\label{S2template}
\end{align}
is over 99\% correlated with $B^{\rm A}_{s=2}$ under the inner product \eqref{inner}.

\bibliographystyle{utphys}
\bibliography{Cross}

\end{document}